\documentclass[12pt]{article}
\usepackage{amsmath,amssymb,graphicx,slashbox}
\numberwithin{equation}{section}

\def\mydate{31 March, 2009}
\def\ignore#1{{}}

\newcounter{sxn}

\newcounter{axn}

\date{}

\newdimen\mybaselineskip
\renewcommand{\baselinestretch}{1.25}

\renewcommand{\thefootnote}{\arabic{footnote}}

\newcommand{\beeq}{\begin{equation}}
\newcommand{\eneq}{\end{equation}}
\newcommand{\beqn}{\begin{eqnarray}}
\newcommand{\eeqn}{\end{eqnarray}}

\def\mybig{\displaystyle \strut }

\def\dd{\partial}
\def\la{\raise.16ex\hbox{$\langle$}\lower.16ex\hbox{}  }
\def\ra{\, \raise.16ex\hbox{$\rangle$}\lower.16ex\hbox{} }
\def\go{\rightarrow}

\def\onehalf{ \hbox{${1\over 2}$} }
\def\half{ {1\over 2} }

\def\tr{{\rm tr \,}}
\def\eff{{\rm eff}}

\def\cL{{\cal L}}
\def\cD{{\cal D}}

\def\EM{{\rm EM}}
\def\diag{{\rm diag ~}}
\def\GUT{{\rm GUT}}
\def\Planck{{\rm Planck}}
\def\KK{{\rm KK}}

\def\ep{\epsilon}
\def\psibar{ \psi \kern-.65em\raise.6em\hbox{$-$} }
\def\psibarl{ \psi \kern-.65em\raise.6em\hbox{$-$} \lower.6em\hbox{} }

\def\Psibar{ {\overline{\Psi}} }

\def\myfrac#1#2{{\mybig #1\over \mybig #2}}

\begin{document}
\thispagestyle{empty}

\baselineskip=12pt

{\small \noindent \mydate ~(corrected)   \hfill OU-HET 605/2008}

{\small \noindent     \hfill RIKEN-TH-131}


\baselineskip=35pt plus 1pt minus 1pt

\vskip 2.5cm

\begin{center}
{\Large \bf Dynamical Electroweak Symmetry Breaking}\\
{\Large \bf in SO(5)$\, \times \,$U(1) Gauge-Higgs Unification}\\
{\Large \bf with Top and Bottom Quarks}\\

\vspace{2.5cm}
\baselineskip=20pt plus 1pt minus 1pt

{\def\thefootnote{\fnsymbol{footnote}}
\bf 
Y.\ Hosotani$^*$,  
K.\ Oda$^*$,
T.\ Ohnuma$^*$
and 
Y.\ Sakamura$^\dagger$ 
}\\
\vspace{.3cm}
{\small $^*${\it Department of Physics, Osaka University,
Toyonaka, Osaka 560-0043, Japan}}\\
{\small $^\dagger${\it RIKEN,
Wako, Saitama 351-0198, Japan}}\\
\end{center}

\vskip 2.0cm
\baselineskip=20pt plus 1pt minus 1pt

\begin{abstract}
An $SO(5)\times U(1)$ gauge-Higgs unification model in the Randall-Sundrum 
warped space with top and bottom quarks is constructed.   Additional  fermions  
on the Planck brane  make exotic particles  heavy by effectively
changing boundary conditions of bulk fermions from 
those determined by  orbifold conditions.
Gauge couplings of a top quark multiplet trigger electroweak symmetry breaking
by the Hosotani mechanism,  simultaneously giving a top quark the observed mass.  
The bottom quark mass is generated by combination of brane interactions and
the Hosotani mechanism, where only one ratio of brane masses is relevant
when the scale of brane masses is much larger than the Kaluza-Klein scale
($\sim 1.5 \,$TeV).
The Higgs mass is predicted to be 49.9 (53.5) GeV for the warp factor 
$10^{15}$ ($10^{17}$).  The Wilson line phase turns out $\onehalf \pi$ and the Higgs
couplings to $W$ and $Z$ vanish so that the LEP2 bound for the Higgs mass is evaded.
In the flat spacetime limit the electroweak symmetry is unbroken.
\end{abstract}


\newpage
\section{Introduction}

The Higgs particle is the only particle  missing in the standard model  
of electroweak interactions.   It is necessary to induce the electroweak
symmetry breaking, and is expected to be discovered at LHC in the near future.
However,  it is not obvious if the Higgs particle appears as described
in the standard model.   Its mass and  couplings to other particles
may deviate from those in the standard model.  

What we are going to learn from LHC experiments is, among others,
the structure and origin of symmetry breaking.  
Both electroweak unified theory and grand unified theory are 
constructed  with higher gauge symmetry, which, in turn,  must break 
down at low energies.   The mechanism of gauge symmetry breaking 
is the backbone of unification.  For the first time in history
we are going to directly see the mechanism of gauge symmetry breaking.
In the current folklore this gauge symmetry breaking is supposed to be triggered 
by a Higgs scalar field.   Unlike gauge interactions, however,  
there seems no guiding principle  to regulate interactions of a Higgs field, 
including its self interactions and Yukawa couplings to fermions.  
Further there arises the gauge hierarchy problem 
when the standard model is implemented in grand unified theory. 

There are many scenarios proposed beyond the standard model.
Besides supersymmetry, the Higgsless scenario and the little Higgs scenario, 
there is another
intriguing scenario called gauge-Higgs unification in which the 4D Higgs 
field is identified with the extra-dimensional component of gauge potentials
in higher dimensional gauge theory
(see e.g.~\cite{review1, TASI1} and references therein).
Symmetry breaking is caused, 
at the quantum level, by dynamics of Wilson line phases in extra
dimensions through the Hosotani mechanism.\cite{YH1, YH2, HHHK}
Higgs couplings are
controlled by gauge principle with a given spacetime background.

Significant progress has been achieved in the gauge-Higgs unification 
scenario in recent years.
Chiral fermions are naturally accommodated on orbifolds.\cite{Pomarol1}
At low energies the standard model is reproduced in 
models based on such groups as $SU(3)$ and 
$SO(5) \times U(1)$.\cite{Antoniadis1}-\cite{Panico1}  
Besides the naturalness of the small Higgs mass may follow from gauge
invariance associated with Wilson line phases.\cite{Lim2}

In models in flat space the Higgs mass is predicted too small, 
typically one order of magnitude smaller than $m_W$, 
and the $WWZ$ coupling may significantly 
deviate from that in the standard model.  
One way out is to construct a model such that the Wilson line phase
takes a sufficiently small value by tuning matter content.\cite{HHKY, Csaki2}
In ref.\ \cite{Csaki2} an $SU(3)$ model has been proposed by incorporating
fermions in {\bf 3}, {\bf 6} and {\bf 10} representations of the group.
Scrucca et al.\ explored a model with localized kinetic terms for gauge fields
to a heavy Higgs field.\cite{ Scrucca}
It has been also discussed that models in six or more dimensions may help Higgs
fields acquire large masses from self couplings.  
Antoniadis et al. discussed a model on $M^4 \times (T^2/Z_2)$.  
The  very existence of Wilson line phases, however, implies flat directions
in the Higgs potential at the tree level, 
resulting a light Higgs field.\cite{Antoniadis1, HNT2}

The alternative way to resolve the difficulties is to consider the 
gauge-Higgs unification 
in the Randall-Sundrum (RS) space.\cite{Pomarol2}-\cite{Csaki3}
It has been argued that the Higgs mass picks up an enhancement factor 
given by the logarithm of the warp factor
in the RS space so that the Higgs mass can fall in the LHC range
for generic values of the Wilson line phase $\theta_H$.\cite{HM}
Despite wave functions of $W$ and $Z$ get rotated in the group space 
at finite $\theta_H$, their profile in the extra dimension remains almost
flat in the RS space, whereas substantial deformation results in flat space.
As a consequence  the $WWZ$ coupling remains universal to high accuracy
irrespective of the value of $\theta_H$ in the RS space unlike the case 
in flat space.\cite{SH1, HS2}    However, the $WWH$ and
$ZZH$ couplings, where $H$ stands for the Higgs field, are suppressed
by a factor $\cos \theta_H$ compared with those in the standard 
model.\cite{SH1, HS2, Giudice1}
This is a distinctive prediction, and can be tested at LHC.
There are also predictions for Yukawa couplings,\cite{HNSS} 
electroweak precision tests,\cite{Panico1, Wagner2, Lim3}  
anomalous magnetic moments\cite{Lim4} etc..

There remains a challenging task to incorporate quarks and leptons 
with dynamical electroweak (EW) symmetry breaking  in the scheme.  
At low energies the quark-lepton content in the standard model
must be reproduced with correct gauge couplings and mass spectrum.  
As a generic feature, fermion multiplets in gauge-Higgs unification tend to
give rise to unwanted exotic light particles, which must be made heavy 
by some means such as brane mass terms.\cite{Nomura1}
More importantly the presence of fermions is vital 
to have EW symmetry breaking.  The fermion content must be
such that it triggers EW symmetry breaking by quantum effects.

To summarize, to include quarks and leptons with dynamical EW symmetry breaking with correct gauge couplings and mass spectra is a highly nontrivial task.
An important step has been put forward 
by Medina, Shar and Wagner who proposed an $SO(5) \times U(1)$ model in the RS space with fermions in {\bf 5} and {\bf 10} representations of $SO(5)$.\cite{Wagner1}
It has been shown that the presence of a top quark induces electroweak symmetry breaking.
The fermion structure of this model is elaborated in order to pass the electroweak precision tests, especially to obtain appropriate $S$ and $T$ parameters as well as the consistent $Zb\bar b$ coupling.
Each generation follows from two {\bf 5} multiplets and one {\bf 10} multiplet in the bulk, where unwanted fields are made heavy by assigning a specific choice of boundary conditions.
These boundary conditions can be achieved, even if one starts from the boundary conditions consistent with orbifolding, by introducing numbers of additional brane fermions in order to effectively modify the original orbifold boundary conditions 
to the desired ones for low-lying modes in the associated Kaluza-Klein towers.
Note that the orbifolding by $Z_2$ parity plays an important role to remove the gravitational instability due to having the negative tension brane on the orbifold fixed point in the Randall-Sundrum spacetime in addition to solving the Einstein equations.
It can be checked that the boundary conditions in ref.~\cite{Wagner1} can be obtained from the orbifold ones by introducing $O(10)$ brane fermion multiplets for each generation.
\ignore{
}
Although the success of the model  is striking, there are many 
free parameters, which make the relation between the parameters in the Lagrangian and the low energy observables obscure.
Therefore it is difficult to transparently see the theoretical structure when one 
further seeks  simpler models.

In the present paper we propose a model with simpler fermion content written
in the form of a Lagrangian with natural boundary conditions dictated 
by the orbifold structure in the RS space as a prototype to clarify the above mentioned relation so that the electroweak symmetry breaking structure is transparent and the low energy fields can be explicitly written in terms of the bulk and brane fields, postponing the full analysis of the $S$ and $T$ parameters.
The model is constructed to fit in with the criteria of ref.\  \cite{Agashe-custodial}
for suppressing radiative corrections to the $\rho$ ($T$) parameter
and $Z b \bar b$ coupling.
Everything should follow from the equations of  motion, or the action principle.  
Brane fermions introduced on the Planck brane, through couplings
with bulk fermions,  effectively change 
boundary conditions of some of fermion components to 
push all unwanted exotic particles to the Kaluza-Klein mass scale.
Quite interestingly the low energy physics does not depend on the values of
various parameters introduced on the Planck brane except for one 
ratio of mass parameters which is fixed by $m_b/m_t$.   Furthermore,
it is found that the effective potential for the Wilson line phase $\theta_H$ is 
minimized at $ \pm \onehalf \pi$ as a consequence of  gauge dynamics of 
heavy top quarks.  At $\theta_H = \pm \onehalf\pi$ 
the EW symmetry is dynamically broken to $U(1)_\EM$ and 
the $WWH$ and $ZZH$ couplings vanish at the tree level.
With $m_W$ and $m_t$ given, the Higgs mass is predicted 
around 50 GeV.  The LEP2 bound for the Higgs mass is evaded
due to the vanishing $ZZH$ coupling.  The model predicts  a light Higgs
particle with a narrow width.

The paper is organized as follows.   The model is specified in Section 2.  
The spectrum in the gauge-Higgs sector, which is known in the literature,
is briefly summarized in Section 3 in the form suited for subsequent applications.
In Section 4 the spectrum of fermions is analyzed by solving coupled equations 
of bulk and brane fermions.  It will be shown how brane mass terms give rise to
discontinuities in bulk fermions at the Planck brane, effectively changing
boundary conditions there.  Equations determining the spectrum take
complicated forms in the presence of brane mass terms.  The expressions are 
tremendously simplified when the scale of brane masses is much larger than
the Kaluza-Klein mass scale $m_\KK$.  
It is found that the top mass $m_t$  is generated 
by the Hosotani mechanism almost independent of brane masses, 
whereas the bottom mass is generated by combination of brane masses
and the Hosotani mechanism.  
With the knowledge of the mass spectrum
the effective potential for the Wilson line phase $V_\eff(\theta_H)$ 
is evaluated to examine
electroweak symmetry breaking in Section 5. 
It will be found that the contribution from a top quark dominates 
over those from the gauge fields such that the potential is minimized 
at $\theta_H = \pm \onehalf \pi$ and the electroweak symmetry is 
dynamically broken.   Contributions from light quarks and leptons are
negligible.   In Section 6 the Higgs mass is evaluated. 
Section 7 is devoted to summary and discussions.

\section{$SO(5)\times U(1)$ model}

The metric in the Randall-Sundrum (RS) warped spacetime\cite{RS1} is given by
\beeq
 ds^2 = G_{MN}dx^M dx^N 
 = e^{-2\sigma (y)} \, \eta_{\mu\nu} dx^\mu dx^\nu
 +dy^2 ~~, 
 \label{metric1}
\eneq
where $\eta_{\mu\nu} = \diag(-1,1,1,1)$, $\sigma (y)=\sigma (y+2L)$, 
and $\sigma (y) \equiv k |y|$ for $|y| \leq  L$.   
The fundamental region in the fifth dimension is given by $0 \le y \le L$,
which is sandwiched by the Planck brane at $y=0$ and the TeV brane at $y=L$,
respectively.  The bulk region $0 <y < L$ is a sliced AdS spacetime with a 
negative cosmological constant $\Lambda = - 6 k^2$.  The metric is specified 
with two parameters, $k$ and $kL$.  In the gauge-Higgs unification scenario
discussed below the $W$ boson mass is predicted as $m_W(k, kL, \theta_H)$
where $\theta_H$ is the Wilson line phase of gauge fields  determined  
dynamically once the matter content is given.  With $m_W$ given, therefore,
there remains only one parameter specifying the spacetime. 
In field theory  in the Randall-Sundrum spacetime there
appear Kaluza-Klein (KK) excitations in a  tower for each particle, 
with the KK mass scale given by
\beeq
m_\KK = \frac{\pi k}{e^{kL} -1} \sim \pi k e^{-kL} ~.
\label{KK1}
\eneq
We shall find that for $e^{kL} = 10^{15}$ ($10^{17}$), 
$k= 4.72 \times 10^{17} \,$GeV
($5.03 \times 10^{19}\,$ GeV) and $m_\KK = 1.48 \,$TeV ($1.58 \,$TeV).
The results in the present paper are insensitive to the value of $k$ in the above 
range.

We consider an $SO(5) \times U(1)_X$ gauge theory in the Randall-Sundrum 
warped spacetime.  The $SO(5) \times U(1)_X$ symmetry is reduced to
$SO(4) \times U(1)_X$ by orbifold boundary conditions.  The symmetry is 
further reduced to $SU(2)_L \times U(1)_Y$ on the Planck brane.
In the present paper
we  address neither a question of how the orbifold structure of spacetime
appears with orbifold conditions,  nor a question of how the symmetry further
reduces to the standard model symmetry  $SU(2)_L \times U(1)_Y$
on the Planck brane.    We imagine these happen
at a high energy scale of $O(M_{\rm GUT})$ to $O(M_{\rm Planck})$ as described below.

The Lagrangian density  consists of four parts.
\beeq
\cL = \cL_{\rm bulk}^{\rm gauge} + \cL_{\rm Pl.\  brane}^{\rm scalar}
+ \cL_{\rm bulk}^{\rm fermion} 
 + \cL_{\rm Pl.\  brane}^{\rm fermion} ~.
\label{Lag1}
\eneq
The bulk parts respect $SO(5) \times U(1)_X$ gauge symmetry.  
There are   $SO(5)$ gauge fields~$A_M$ and $U(1)_X$ gauge field~$B_M$. 
The former are decomposed as 
$ A_M = \sum_{a_L =1}^3 A^{a_L}_M T^{a_L} 
+\sum_{a_R =1}^3 A^{a_R}_M T^{a_R}
 +\sum_{\hat{a}=1}^4 A^{\hat{a}}_M T^{\hat{a}}$
 where  $T^{a_L, a_R}$ ($a_L, a_R =1,2,3$) and $T^{\hat{a}}$ 
($\hat{a}=1,2,3,4$) are the generators of $SO(4)\sim SU(2)_L \times SU(2)_R$ 
and $SO(5)/SO(4)$, respectively.  $\cL_{\rm bulk}^{\rm gauge}$ is given by
\beeq
\cL_{\rm bulk}^{\rm gauge} = 
 -\tr  \frac{1}{2} F^{(A)MN}F^{(A)}_{MN} 
 - \frac{1}{4} F^{(B)MN}F^{(B)}_{MN}
 \label{Lag2}
\eneq
with the associated gauge fixing and ghost terms, 
where $ F^{(A)}_{MN} = \dd_M A_N -\dd_N A_M -ig_A[A_M,A_N] $ and
$F^{(B)}_{MN} = \dd_M B_N -\dd_N B_M$.  

The orbifold boundary conditions at $y_0=0$ and $y_1=L$ for gauge fields 
are given by
\beqn
&& \hskip -1cm
 \begin{pmatrix} A_\mu \cr A_y \end{pmatrix} (x, y_j-y) 
= P_j  \begin{pmatrix} A_\mu \cr - A_y \end{pmatrix}  (x,y_j +y) P_j^{-1} 
 ~~, \cr
\noalign{\kern 10pt}
&& \hskip -1cm  
\begin{pmatrix} B_\mu \cr B_y \end{pmatrix} (x, y_j-y) 
= \begin{pmatrix} B_\mu \cr - B_y \end{pmatrix} (x,y_j+ y)  ~~, \nonumber\\
\noalign{\kern 10pt}
&& \hskip -0.5cm
P_j = \hbox{diag} \, (-1, -1, -1, -1, +1) ~~~ (j= 0, 1) ~,
\label{BC1}
\eeqn
which reduce the $SO(5) \times U(1)_X$ symmetry to $SO(4) \times U(1)_X$.
We introduce a scalar field $\Phi (x)$ on the Planck brane which belongs to
$(0, \onehalf )$ representation of $SO(4) = SU(2)_L \times SU(2)_R$ and
has a charge  of $U(1)_X$.  With
\beqn
&&\hskip -1.cm
 \cL_{\rm Pl.\  brane}^{\rm scalar} =
 \delta (y) \Big\{ - (D_\mu \Phi)^\dagger D^\mu \Phi 
 - \lambda_\Phi  (\Phi^\dagger \Phi - w^2)^2 \Big\} ~,  \cr
 \noalign{\kern 10pt}
 &&\hskip -1cm
 D_\mu \Phi = \dd_\mu \Phi 
  - i \Big( g_A  \sum_{{a_R}=1}^3 A^{a_R}_\mu T^{a_R}  
  + \frac{g_B }{2} B_\mu \Big)  \Phi
 \label{Lag3}
 \eeqn
the $SU(2)_R \times U(1)_X$
symmetry breaks down to $U(1)_Y$,  the massless modes of 
$A_\mu^{1_R}$, $A_\mu^{2_R}$, and $A_\mu^{\prime 3_R}$ acquiring 
large masses.  Here 
\beqn
&&\hskip -1cm
\begin{pmatrix} A^{\prime 3_R}_M \cr B^Y_M \end{pmatrix} =
\begin{pmatrix} c_\phi & - s_\phi \cr s_\phi & c_\phi \end{pmatrix}
\begin{pmatrix} A^{ 3_R}_M \cr B_M \end{pmatrix}  ~, \cr
\noalign{\kern 10pt}
&&\hskip -1cm
c_\phi  \equiv \frac{g_A}{\sqrt{g_A^2+g_B^2}} ~~,~~
s_\phi  \equiv \frac{g_B}{\sqrt{g_A^2+g_B^2}} ~~.  
\label{planck1}
\eeqn
We suppose that $w$ is much bigger than the Kaluza-Klein mass scale, 
being of $O(M_{\rm GUT})$ to $O(M_{\rm Planck})$.
The net effect for low-lying modes of the Kaluza-Klein towers of 
$A_\mu^{1_R}$, $A_\mu^{2_R}$, and $A_\mu^{\prime 3_R}$ is that they
effectively obey Dirichlet boundary conditions at the Planck brane.
We note that the effective change of boundary conditions occurs for $A_\mu$ 
components, but not for $A_y$ components as seen from (\ref{Lag3}).
This  also conforms with the invariance under large gauge transformations
which shift the Wilson line phase by a multiple 
of $2\pi$.
\ignore{
\footnote{The boundary conditions adopted in ref.\ \cite{Wagner1} 
are the same for the $A_\mu$ components as ours, but slightly differ from ours 
for the $A_y$ components which obey normal orbifold boundary conditions.
See the resultant boundary conditions tabulated in Table I of ref.\ \cite{HS2}.}
}
We see in the subsequent sections, in a concrete manner,   that a similar 
effective change of boundary conditions takes place for fermions as well.

We remark that massive modes in the  Kaluza-Klein towers of the $A_y$
components are unphysical.   Their spectrum, in general, depends on 
gauge-fixing conditions imposed.   In the bulk we adopt the gauge-fixing 
in refs. \cite{Oda1, HNSS} so that the kinetic terms of the $A_\mu$ and 
$A_y$ of $SO(5) \times U(1)_X$ become diagonal.

On the Planck brane at $y=0$ one further adds a gauge-fixing
condition ($\propto \delta(y)$) suitable for the gauge symmetry breaking 
$SU(2)_R \times U(1)_X  \go U(1)_Y$ induced by the scalar field $\Phi(x)$.
The most convenient choice is the standard $R_\xi$ gauge, in which 
the ghost fields in the bulk  have the same boundary conditions and 
spectrum as the corresponding $A_\mu$ components for $w \gg M_\KK$.

In our scheme the $A_y$ components of $SU(2)_R \times U(1)_X$ are
odd under parity around $y=0$ and obey the Dirichlet condition so that
they do not enter into the gauge-fixing conditions at the Planck brane.
It is possible to allow discontinuities in $A_y$ at $y=0$ by enlarging 
the configuration space for $A_y$.  In this case one may include 
discontinuities in $A_y$ in the gauge-fixing condition at the Planck brane.
This possibility is examined in ref.\ \cite{TASI1}.  It is shown there the 
boundary conditions for $A_y$ at $y=0$ and the resulting spectrum
depend on the gauge parameters introduced.  In one limit (the $\xi$ 
parameter in the bulk approaching 0) 
$A_y$ obeys the Dirichlet boundary conditions at $y=0$, 
whereas in another limit ($\xi \go \infty$) corresponding to the unitary gauge
$A_y$ obeys the Neumann boundary conditions which were adopted 
in ref.\ \cite{Wagner1}.  In this paper all $A_y$ components are supposed to
be continuous at the Planck and TeV branes.

The resultant spectrum at low energies in the gauge sector is that of the 
standard model as discussed in detail in ref.\ \cite{HS2}.  
There appear $W$, $Z$ bosons and photons as 
massless gauge fields, and an $SU(2)_L$ doublet Higgs boson $\phi$ from the 
$A_y$ component.  The Higgs boson, which is nothing but  a fluctuation mode 
of the Wilson line phase $\theta_H$,  is massless at the tree level, 
but becomes massive at the quantum level.  We shall see below that
the effective potential  $V_\eff(\theta_H)$ is minimized at
$\theta_H = \pm \onehalf \pi$ due to a contribution from the top quark 
multiplet so that
the electroweak symmetry breaks down to $U(1)_{\rm EM}$.  At the same
time the Higgs field acquires a mass $\sim 50\,$GeV.

To describe top and bottom quarks we introduce two fermion multiplets 
in the vector representation of $SO(5)$.  In the bulk one has
\beqn
&&\hskip -1.cm 
\cL_{\rm bulk}^{\rm fermion} = 
\sum_{a=1}^2 \Psibar_a \cD (c_a) \Psi_a ~, \cr
\noalign{\kern 5pt}
&&\hskip -1.cm 
\cD (c_a) =  \Gamma^A {e_A}^M \Big(
\dd_M + \frac{1}{8} \omega_{MBC} [\Gamma^B, \Gamma^C] 
 - ig_A A_M -i q_a \frac{g_B}{2}  B_M \Big) -  c_a \sigma'(y)  ~.
\label{Lag4}
\eeqn
Here ${e_A}^M$'s are tetrads and $\Gamma^A$'s are Dirac matrices in the 
orthonormal frame.  We take in the present paper 
\beqn
&&\hskip -1cm 
\Gamma^\mu = 
\begin{pmatrix} & \sigma^\mu \cr \bar \sigma^\mu \end{pmatrix}
~~(\mu=0,1,2, 3) ~~,~~
\Gamma^5 = 
\begin{pmatrix} 1 \cr& -1 \end{pmatrix} ~~, \cr
\noalign{\kern 10pt}
&&\hskip -1cm 
\sigma^\mu = (1, \vec \sigma) ~~,~~ \bar \sigma^\mu = (-1, \vec \sigma) ~~.
\label{DiracM}
\eeqn
The $c_a$ term in (\ref{Lag4}) gives a bulk kink mass,\cite{GP}  where
$\sigma'(y) = k \, \ep (y)$ is a periodic step function with a magnitude $k$.
The dimensionless  parameter $c_a$ plays an important role in the Randall-Sundrum
warped spacetime.  The $U(1)_X$ charges are $q_1 = \frac{2}{3}$ and
$q_2 = - \frac{1}{3}$.  The notation  $\Psibar = i \Psi^\dagger \Gamma^0$
has been adopted.  The orbifold boundary conditions are given by
\beeq
\Psi_a(x, y_j - y) =  P_j \Gamma^5  \Psi_a (x, y_j + y) ~~.
\label{BC2}
\eneq
With $P_j$ in (\ref{BC1}) the first four components of $\Psi_a$ are even 
under parity for the 4D left-handed ($\Gamma^5 = -1$) components.

To facilitate discussions below,  it is useful to express the 
$SU(2)_L \times SU(2)_R$ content of an $SO(5)$  vector $\Psi$.
The first four components $\psi_k$ ($k=1, \cdots, 4$) belong to
$( \onehalf ,  \onehalf)$, whereas the fifth component $\psi_5$ 
to $(0,0)$.  
We define $\hat \psi$ by
\beqn
\hat{\psi} &=& 
\begin{pmatrix} \hat{\psi}_{11} & \hat{\psi}_{12} \cr 
      \hat{\psi}_{21} & \hat{\psi}_{22} 
\end{pmatrix}  \cr
\noalign{\kern 5pt}
&\equiv& \frac{1}{\sqrt{2}}(\psi_4+i\vec{\psi}\cdot\vec{\tau}) i\tau_2
=  \frac{1}{\sqrt{2}} 
\begin{pmatrix} -i \psi_1 - \psi_2 & i \psi_3 + \psi_4 \cr 
     i \psi_3 - \psi_4 & i \psi_1 - \psi_2
\end{pmatrix}
~~, 
\label{def1}
\eeqn
which transforms, under an $SU(2)_L \times SU(2)_R$ transformation 
$\Omega_L \otimes \Omega_R$, as 
$\hat \psi' = \Omega_L \, \hat \psi  \, {\Omega_R}^t$.
$(\hat{\psi}_{11},\hat{\psi}_{21})^t$ and 
$(\hat{\psi}_{12},\hat{\psi}_{22})^t$ form 
$SU(2)_L$ doublets. 
In the following we express components of $\Psi$ as 
$\Psi = (\hat{\psi}_{11}, \hat{\psi}_{21}, \hat{\psi}_{12}, 
\hat{\psi}_{22}, \psi_5)^t$. 
With this notation, the two fermion multiplets contain
\beeq
\Psi_1 = 
\begin{pmatrix} T \cr B \cr t \cr b \cr  t' \end{pmatrix}
~~
\begin{matrix} \frac{5}{3} \cr \frac{2}{3} \cr \frac{2}{3}  \cr
      -\frac{1}{3} \cr  \frac{2}{3} \end{matrix}
~~~,~~~
\Psi_2 = 
\begin{pmatrix} U \cr D \cr X \cr Y \cr  b' \end{pmatrix} 
~~
\begin{matrix} \frac{2}{3} \cr -\frac{1}{3} \cr -\frac{1}{3}  \cr
      -\frac{4}{3} \cr  -\frac{1}{3} \end{matrix}
~~~.
\label{fermion1}
\eneq
The numbers written on the side are  values of the electric charge 
 $Q_E=T^{3_L}+T^{3_R}+Q_X$. 
$(T,B)$, $(t,b)$, $(U,D)$ and $(X,Y)$ form $SU(2)_L$ doublets.
$\psi_4$, which couples to $\psi_5$ through the Wilson line phase
$\theta_H$, is given by $(\hat{\psi}_{12} - \hat{\psi}_{21})/\sqrt{2}$.

\ignore{
$\psi_4$ is $(t - B)/\sqrt{2}$ and  $(X-D)/\sqrt{2}$ for
$\Psi_1$ and $\Psi_2$, respectively.  
}

If there were no additional interactions on the branes,  there would appear,
before the electroweak (EW) symmetry breaking, massless modes in 4D
in $SU(2)_L$ multiplets 
\beqn
&&\hskip -1cm
Q_{1L}= \begin{pmatrix} T_L \cr B_L  \end{pmatrix} ~,~
~ q_L = \begin{pmatrix} t_L \cr b_L  \end{pmatrix} ~ ~,~
~  t_R'  ~,~ \cr
 \noalign{\kern 10pt}
 &&\hskip -1cm
Q_{2L} = \begin{pmatrix} U_L \cr D_L  \end{pmatrix} ~,~
Q_{3L} = \begin{pmatrix} X_L \cr Y_L\end{pmatrix} ~,~
  b_R' ~.
\label{fermion2}
\eeqn
After the EW symmetry breaking by the Hosotani mechanism the top  
quark and extra $b'$ would acquire  finite masses, but
  other left-handed  modes would remain massless.
We want the top and bottom to acquire the observed masses, whereas
other light modes to gain large masses of  the Kaluza-Klein scale.  

We show in the present paper that  the presence of brane fermions~\cite{Nomura1} on the
Planck brane (at $y=0$) cures these problems naturally.   
We introduce three right-handed multiplets belonging to $(\onehalf, 0)$ 
representation of $SU(2)_L \times SU(2)_R$;
\beeq
\hat \chi_{1R} = \begin{pmatrix} \hat T_R \cr \hat B_R  \end{pmatrix}_{7/6} ~,~
\hat \chi_{2R} = \begin{pmatrix} \hat U_R \cr \hat D_R  \end{pmatrix}_{1/6} ~,~
\hat \chi_{3R} = \begin{pmatrix} \hat X_R \cr \hat Y_R  \end{pmatrix}_{-5/6} ~.
\label{fermion3}
\eneq
Here the subscripts $7/6$ etc.\ represent the $U(1)_X$ charges of
the corresponding multiplets.  We write down a general $SU(2)_L \times U(1)_Y$
invariant brane action;
\beqn
&&\hskip -1.cm
\cL_{\rm Pl.\  brane}^{\rm fermion} = i \delta (y)   \bigg\{
\sum_{\alpha =1}^3 
\hat \chi_{\alpha R}^\dagger  \, \bar \sigma^\mu  D_\mu  \,  \chi_{\alpha R}  \cr
\noalign{\kern 10pt}
&&\hskip -0.5cm
- \sum_{\alpha =1}^3  \mu_\alpha 
\big( \hat \chi_{\alpha R}^\dagger Q_{\alpha L}
- Q_{\alpha L}^\dagger  \hat \chi_{\alpha R} \big)
- \tilde \mu \big( \hat \chi_{2 R}^\dagger \, q_L
- q_L^\dagger  \,  \hat \chi_{2 R} \big)
 ~ \bigg\}
 \label{Lag5}
\eeqn
where $D_\mu$ in the kinetic term has the same form as in (\ref{Lag3})
with $A_\mu^{a_R} T^{a_R}$ replaced by $A_\mu^{a_L} T^{a_L}$. 
There are four brane mass parameters, $\mu_\alpha$ and $\tilde \mu$, 
which have dimensions of (mass)$^{1/2}$.  
In the subsequent discussions we suppose that $\mu_\alpha^2$ and
$\tilde \mu^2$ are much larger than the Kaluza-Klein scale $m_\KK \sim 1.5 \,$TeV,
possibly being of order $M_\GUT$ or $M_\Planck$.  It will be shown  below
that the only relevant parameter for the spectrum at low energies is 
the ratio $\tilde \mu/\mu_2$ so long as $\mu_\alpha^2, \tilde \mu^2 \gg m_\KK$.

In ref.\ \cite{Wagner1} a model with top and bottom quarks residing  in two {\bf 5}
multiplets  and one {\bf 10} multiplet has been considered.  It is assumed 
that the fermions satisfy boundary conditions differing from those obtained by
simple orbifolding.   It has been stated there that this change of boundary conditions 
can follow from brane mass interactions at the TeV brane.   We show below by solving 
equations of motion that
desired change of boundary conditions for low-lying modes in Kaluza-Klein towers
takes place as a result of brane mass interactions at the Planck brane (\ref{Lag5}), 
keeping  the custodial $SO(4)$ symmetry at the TeV brane.

\section{Spectrum in the gauge-Higgs sector}

The spectrum in the gauge-Higgs sector described by 
$\cL_{\rm bulk}^{\rm gauge} + \cL_{\rm Pl.\  brane}^{\rm scalar}$ with 
(\ref{Lag2}) and (\ref{Lag3}) has been well spelled out in ref.\ \cite{HS2}.
In this section we summarize the results obtained there, 
which becomes necessary to evaluate the effective potential for the Wilson line 
phase $\theta_H$.  

The $y$ coordinate in the Randall-Sundrum spacetime in (\ref{metric1}) is 
suited for seeing the orbifold structure.  In finding the spectrum of particles 
and their wave functions in the fifth dimension the conformal coordinate
$z \equiv e^{\sigma(y)}$ is useful, with which the metric becomes
\beeq
 ds^2 = \frac{1}{z^2} \bigg\{ \eta_{\mu\nu} dx^\mu dx^\nu
     +\frac{dz^2}{k^2}  \bigg\} ~~.
\label{metric2}
\eneq
The fundamental region $0\le y\le L$ is mapped to $1 \le z \le z_L = e^{kL}$. 
$z_L$ is called as a warp factor, which we will 
find to be around $10^{15}$ to $10^{17}$.
In the bulk region $0 < y < L$, one has 
$\dd_y = kz \dd_z$, $A_y = kz A_z$ etc. 

The fifth dimensional component of gauge potentials $A_y$ or $A_z$ has 
zero modes in the $SO(5)/SO(4)$ part with generators $T^{\hat a}$
($\hat a = 1, \cdots, 4$); 
\beqn
&&\hskip -1cm
 A_z^{\hat a} (x, z)   = \phi^a (x) \sqrt{ \frac{2}{k (z_L^2 -1)} } \, z + \cdots ~, \cr
\noalign{\kern 10pt}
&&\hskip -1cm
\Phi_H (x) = \frac{1}{\sqrt{2}} 
\begin{pmatrix} \phi^2 + i \phi^1 \cr \phi^4 - i \phi^3 \end{pmatrix} ~. 
\label{Higgs1}
\eeqn
$\Phi_H$ corresponds to the $SU(2)_L$ doublet Higgs field in the standard model.
Without loss of generality one can assume that $\la \phi^a \ra = v \, \delta^{a4}$
when the EW symmetry is spontaneously broken.  
The Wilson line phase $\theta_H$ is given by
$\exp \big\{ \frac{i}{2} \theta_H (2 \sqrt{2} \,  T^{\hat 4} ) \big\} =
\exp \big\{ ig_A  \int_1^{z_L} dz \, \la A_z \ra \big\}$ so that \cite{HM} 
\beeq
\theta_H = \frac{1}{2} g_A v \sqrt{ \frac{z_L^2 - 1}{k} }
\sim \frac{g_4 v}{2} \, \frac{\pi \sqrt{kL}}{m_\KK}   ~~.
\label{Wilson1}
\eneq
Here $g_4 = g_A/ \sqrt{L}$ is the four-dimensional $SU(2)_L$
gauge coupling constant.  We remark that $\theta_H$ is a phase variable so that
physics is periodic in $\theta_H$ with a period $2\pi$. 

The spectrum is determined with $\theta_H \not= 0$.  Various components 
in $SO(5)$ mix among each other.  
Following Falkowski,\cite{Falkowski1} 
we define basis functions for mass eigenmodes 
in the gauge-Higgs sector  by solutions of the Bessel equation
\beqn
&&\hskip -1cm
\Big( \frac{d^2}{dz^2} - \frac{1}{z} \frac{d}{dz} + \lambda^2 \Big)
\begin{pmatrix} C(z; \lambda)  \cr S(z; \lambda) \end{pmatrix} = 0 ~~, \cr
\noalign{\kern 10pt}
&&\hskip -.5cm
C(z_L ; \lambda) = z_L ~~,~~  C' (z_L ; \lambda) = 0 ~~, \cr
\noalign{\kern 5pt}
&&\hskip -.5cm
S(z_L ; \lambda) = 0 ~~~,~~   S' (z_L ; \lambda) = \lambda ~~.
\label{Bessel1}
\eeqn
Here $C' = dC/dz$  and a relation $C S' - S C' = \lambda z$ holds.
The Neumann and Dirichlet boundary conditions for $A_\mu$ correspond to
solutions $C(z_L ; \lambda)$ and $S(z_L ; \lambda)$, respectively.
The dimensionless eigenvalue $\lambda$ is related to a 4D mass by $m = k\lambda$. 
As the scalar interactions on the Planck brane at $z=1$ effectively change
the boundary conditions there, it is most convenient to use the base
functions $C$ and $S$ defined with boundary conditions at $z=z_L$
as in (\ref{Bessel1}).  They generalize trigonometric functions in
flat space, and are given in terms of Bessel functions by
\beqn
&&\hskip -1cm
C(z; \lambda) = \frac{\pi}{2} \lambda z z_L \, F_{1,0} (\lambda z, \lambda z_L) 
~~,~~
C'(z; \lambda) = \frac{\pi}{2} \lambda^2 z z_L \, 
       F_{0,0} (\lambda z, \lambda z_L) ~, \cr
\noalign{\kern 10pt}
&&\hskip -1cm
S(z; \lambda) = - \frac{\pi}{2} \lambda z \, F_{1,1} (\lambda z, \lambda z_L) 
~~,~~
S'(z; \lambda) = - \frac{\pi}{2} \lambda^2 z  \, F_{0,1} (\lambda z, \lambda z_L) ~.
\label{Bessel2}
\eeqn
Here $F_{\alpha,\beta}(u,v)$ is defined as
\beeq
F_{\alpha,\beta}(u,v) = J_\alpha (u) Y_\beta (v) -  Y_\alpha (u) J_\beta (v)
\label{Bessel3}
\eneq
which is the same as in  ref.\ \cite{HS2}.  
The relation $F_{\alpha, \alpha-1} (u,u)=2/\pi u$ has been used in (\ref{Bessel2}).

\subsection{KK towers of 4D gauge fields}

\noindent
(i) $(1_L, 1_R, \hat 1), (2_L, 2_R, \hat 2)$ components ($W$-tower)

With $\theta_H \not= 0$, three components 
$A_\mu^{a_L}$,  $A_\mu^{a_R}$, and $A_\mu^{\hat a}$ ($a= 1,2$) 
mix among each other.  The mass spectrum $m_n = k \lambda_n$ is determined by
\beqn
&&\hskip -1cm
C(1; \lambda_n) = 0 ~~, 
\label{gHspec1a} \\
\noalign{\kern 10pt}
&&\hskip -1cm
2 S(1; \lambda_n) C'(1; \lambda_n)  + \lambda_n \sin^2 \theta_H = 0  \quad
\hbox{($W$ tower)}.
\label{gHspec1b}
\eeqn
The spectrum (\ref{gHspec1a}) contains only massive modes.
The spectrum (\ref{gHspec1b}), which depends on $\theta_H$,  
contains a $W$ boson as the lowest mode $\lambda_0$.   
When the warp factor $z_L = e^{kL} \gg 1$,  one finds that 
$\lambda_0 z_L \ll 1$ for any value of $\theta_H$.  
Employing approximate formulas for
Bessel functions, one finds that the $W$ boson mass is given by
\beeq
m_W \sim  \sqrt{\frac{k}{L}} \, e^{-kL}  \, | \sin\theta_H | 
\sim  \frac{m_\KK}{\pi \sqrt{kL} } \, | \sin\theta_H |  ~~.
\label{Wmass}
\eneq
Later it will be found that the effective potential is minimized at $\theta_H = \onehalf \pi$.
Hence, we find that for $e^{kL} = 10^{15}$ ($10^{17}$), 
$k= 4.72 \times 10^{17} \,$GeV
($5.03 \times 10^{19}\,$ GeV) and $m_\KK = 1.48 \,$TeV ($1.58 \,$TeV).

\vskip 5pt

\noindent
(ii) $(3_L, 3_R, \hat 3, B)$  or $(3_L, 3_R', \hat 3, Y)$ components 
($\gamma$- and $Z$-towers)

Four components $A_\mu^{3_L}$,  $A_\mu^{3_R}$,  $A_\mu^{\hat 3}$, 
and $B_\mu$  mix among each other.  The spectrum is given by
\beqn
&&\hskip -1cm
C' (1; \lambda_n) = 0 \quad \hbox{(photon tower)}, 
\label{gHspec2a} \\
\noalign{\kern 10pt}
&&\hskip -1cm
C(1; \lambda_n) = 0 ~~, 
\label{gHspec2b} \\
\noalign{\kern 10pt}
&&\hskip -1cm
2 S(1; \lambda_n) C'(1; \lambda_n)  
+ \lambda_n ( 1 + s_\phi^2) \sin^2 \theta_H = 0 \quad
\hbox{($Z$ tower)}.
\label{gHspec2c}
\eeqn
Here $s_\phi$ is defined in (\ref{planck1}).  
The spectrum  (\ref{gHspec2a}) contains a zero mode $\lambda_0=0$, 
corresponding to a photon.  The spectrum  (\ref{gHspec2c}) contains a $Z$ boson,
whose mass is  given by
\beeq
m_Z \sim \frac{m_W}{\cos \theta_W} ~~~,~~~
\cos \theta_W = \frac{1}{\sqrt{1 + s_\phi^2}} ~~.
\label{Zmass}
\eneq
The approximate equality is valid to the $O(0.1\%)$ accuracy 
for $m_{KK}=O ({\rm TeV})$.\cite{HS2} 
Notice that the Weinberg angle $\theta_W$ is almost independent of $\theta_H$,
which is not the case in  the corresponding model in flat spacetime 
$M^4 \times (S^1/Z_2)$.

We would like to add a comment on wave functions.  
The profile of the photon wave function
in the fifth dimension is exactly constant.  
The $W$ and $Z$ wave functions are almost constant in the fifth coordinate 
except for the vicinity of the TeV brane, though they have significant $\theta_H$ dependence
in the weight of the $SO(5)$ group components.  It has been known that the approximate
flatness in the fifth dimension assures the universality in the $WWZ$, $WWWW$ and 
$WWZZ$ couplings, whereas the nontrivial $\theta_H$ dependence in the group
space leads to the deviation of $WWH$ and $ZZH$ couplings from those in the 
standard model.\cite{SH1, HS2, Sakamura1}
In the flat space the $W$ and $Z$ wave functions acquire significant
dependence on  the fifth coordinate when $\theta_H$ becomes $O(1)$,
which leads to the deviation of $WWZ$ coupling from that in the standard model.
\cite{HS2}

\vskip 5pt

\noindent
(iii) $(\hat 4)$  component

The spectrum of $A_\mu^{\hat 4}$ is given by $C(1; \lambda_n) = 0$.
\ignore{
\beeq
C(1; \lambda_n) = 0 ~~.
\label{gHspec3}
\eneq
}
It contains only massive modes.

\subsection{KK towers of 4D scalar fields}

Mode functions of the extra-dimensional component $A_z$ satisfy,
in place of (\ref{Bessel1}), 
\beeq
\Big( \frac{d^2}{dz^2} - \frac{1}{z} \frac{d}{dz} + \frac{1}{z^2} + \lambda^2 \Big)
\begin{pmatrix} C'(z; \lambda)  \cr S'(z; \lambda) \end{pmatrix}  = 0 ~~.
\label{Bessel4}
\eneq
The Neumann and Dirichlet boundary conditions for $A_z$ correspond to
solutions $S'(z_L ; \lambda)$ and $C'(z_L ; \lambda)$, respectively.
In classifying the spectra for $A_z$ it is convenient to introduce
\beeq
\begin{pmatrix} A_z^{a_V} \cr A_z^{a_A} \end{pmatrix}
= \frac{1}{\sqrt{2}} 
\begin{pmatrix} A_z^{a_L} +  A_z^{a_R} \cr
                          A_z^{a_L} -  A_z^{a_R}   \end{pmatrix}
  \quad (a=1,2,3) ~.
\label{defVA}
\eneq

\vskip 5pt

\noindent
(i) $(1_V, 2_V, 3_V, B)$  components 

The spectrum is determined by $C(1; \lambda_n) = 0$, 
\ignore{
\beeq
v ~~,
\label{scalarspec1}
\eneq
}
which contains no zero mode.

\vskip 5pt
\noindent
(ii) $(a_A, \hat a)$ ($a=1,2,3$) components 

The spectrum is given by
\beeq
S(1; \lambda_n) C'(1; \lambda_n)  + \lambda_n \sin^2 \theta_H = 0 ~~.
\label{scalarspec2}
\eneq
We note that this spectrum is obtained for $A_z$ satisfying the orbifold
boundary conditions which are not modified by the additional dynamics
on the Planck brane described by (\ref{Lag3}).  As described in section 2 and 
ref.\ \cite{HS2},
it is related to the large gauge invariance.  The spectrum for this part 
is different from that used in ref.\ \cite{Wagner1}.

\vskip 5pt
\noindent
(iii) $(\hat 4)$ component (Higgs tower)

The spectrum is given by $\lambda_n S(1; \lambda_n) = 0$.
\ignore{
\beeq
\lambda_n S(1; \lambda_n) = 0 ~~.
\label{scalarspec3}
\eneq
}
There is a zero mode $\lambda_0=0$, corresponding to the physical neutral
Higgs field $\phi^4$ in (\ref{Higgs1}).  It acquires a finite mass quantum mechanically
by the Hosotani mechanism.

\subsection{KK towers of ghost fields}

The free part of the equations obeyed by the ghost fields in the bulk are
the same as for the $A_\mu$ part.  
The boundary conditions obeyed by the ghost fields for the group components 
outside $SU(2)_R \times U(1)_X / U(1)_Y$ are obviously the same as for
the $A_\mu$.   
Even for the  group components in $SU(2)_R \times U(1)_X/ U(1)_Y$, 
as explained in Section 2, 
the ghost fields obey the same boundary conditions at the Planck brane as $A_\mu$,
once the $R_\xi$ gauge is adopted on the Planck brane.  
Hence in this gauge all components of the ghost fields  have the same spectrum 
as the corresponding $A_\mu$.

\section{Spectrum of fermions}

The fermion spectrum is found in a similar manner.  The presence of boundary
interactions on the Planck brane (\ref{Lag5}) among bulk fermions $\Psi_a$ 
and brane fermions $\hat \chi_{jR}$ induces discontinuities in a part of the bulk
fermion fields.  It also effectively changes boundary conditions at the Planck 
brane, yielding a desired mass spectrum.
The role of brane mass terms for making exotic fermions heavy  was 
discussed by Burdman and Nomura several years ago.\cite{Nomura1}
We shall see below how this is achieved by solving equations of motion
for both bulk and brane fermions.

\subsection{Basis functions}

Before writing down full equations in the presence of a non-vanishing Wilson line
phase $\theta_H$, let us recall the basic structure of Dirac equations in the absence of
gauge interactions.  In the Randall-Sundrum spacetime the rescaled 
spinor field in the bulk, $\check \Psi = e^{-2\sigma} \Psi = z^{-2} \Psi$, satisfies
a simple equation.  If there were no brane interactions, it would obey
\beqn
&&\hskip -1cm
\left\{ \begin{pmatrix} & \sigma \dd \cr \bar \sigma \dd \end{pmatrix} 
- k \begin{pmatrix} D_- (c) \cr & D_+ (c) \end{pmatrix}  \right\}
\begin{pmatrix} \check \Psi_R \cr \check \Psi_L \end{pmatrix} = 0 ~~, 
\label{Feq1} \\
\noalign{\kern 10pt}
&&\hskip -0.cm
D_\pm (c) = \pm \frac{d}{dz} + \frac{c}{z} ~~.
\label{Feq2}
\eeqn
Here $c$ is the bulk mass parameter, and $\Psi_R$ ($\Psi_L$) represents
right-handed component with $\Gamma^5=+1$ (left-handed component
with $\Gamma^5=-1$).   The parity of $\Psi_R$ at the brane is opposite 
to that of $\Psi_L$.  Without brane interactions the component with even (odd) 
parity satisfies a Neumann (Dirichlet) condition there.  Neumann 
conditions for $\check \Psi_R$ and $\check \Psi_L$ are given by 
$D_-(c) \check \Psi_R=0$ and  $D_+(c) \check \Psi_L=0$, respectively.  

In the rest of the present paper
we always discuss fermions in terms of rescaled fields $\check \Psi$.
To simplify expressions we henceforth drop a symbol check ($\check ~$).
Repeated use of the equation (\ref{Feq1}) gives
\beqn
&&\hskip -1cm
\big\{ \dd^2 - k^2 D_+(c) D_-(c) \big\} \Psi_R = 0 ~~, \cr
\noalign{\kern 5pt}
&&\hskip -1cm
\big\{ \dd^2 - k^2 D_-(c) D_+(c) \big\} \Psi_L = 0 ~~.
\label{Feq3}
\eeqn
Taking into account the fact 
that there are brane interactions at the Planck brane at $z=1$, we
define basis functions for fermions as
\beqn
&&\hskip -1cm
\big\{ D_+(c) D_-(c)  - \lambda^2 \big\} 
\begin{pmatrix} C_R(z; \lambda, c) \cr S_R(z; \lambda, c) \end{pmatrix} = 0 ~~, \cr
\noalign{\kern 10pt}
&&\hskip -1.cm
\big\{ D_-(c) D_+(c)  - \lambda^2 \big\} 
\begin{pmatrix} C_L(z; \lambda, c) \cr S_L(z; \lambda, c) \end{pmatrix} = 0 ~~, \cr
\noalign{\kern 10pt}
&&\hskip -1.cm
C_R= C_L = 1 ~~,~~ D_-(c) C_R = D_+(c) C_L =0 ~~, \cr
\noalign{\kern 5pt}
&&\hskip -1.cm
S_R = S_L = 0 ~~,~~ D_-(c) S_R = D_+(c) S_L =\lambda  \hskip 1cm \hbox{at} ~ z= z_L.
\label{Bessel5}
\eeqn
Explicit forms of these functions are given by
\beqn
&&\hskip -1cm
C_L(z; \lambda, c) = + \frac{\pi}{2} \lambda \sqrt{z z_L}
        \, F_{c+\onehalf, c-\onehalf} (\lambda z, \lambda z_L) ~, \cr
\noalign{\kern 10pt}
&&\hskip -1cm
S_L (z; \lambda, c) = -\frac{\pi}{2} \lambda  \sqrt{z z_L}  \, 
       F_{c+\onehalf, c+\onehalf} (\lambda z, \lambda z_L) ~, \cr
\noalign{\kern 10pt}
&&\hskip -1cm
C_R(z; \lambda, c) = - \frac{\pi}{2} \lambda \sqrt{z z_L}
        \, F_{c-\onehalf, c+\onehalf} (\lambda z, \lambda z_L) ~, \cr
\noalign{\kern 10pt}
&&\hskip -1cm
S_R (z; \lambda, c) = + \frac{\pi}{2} \lambda  \sqrt{z z_L}  \, 
       F_{c-\onehalf, c-\onehalf} (\lambda z, \lambda z_L) ~,
\label{Bessel6}
\eeqn
where $F_{\alpha, \beta}$ is defined in (\ref{Bessel3}). 
These functions are related to each other by
\beqn
&&\hskip -1cm
D_+(c) \begin{pmatrix} C_L \cr S_L \end{pmatrix} =
\lambda \begin{pmatrix} S_R \cr C_R \end{pmatrix}  ~~,~~
D_-(c) \begin{pmatrix} C_R \cr S_R \end{pmatrix} =
\lambda \begin{pmatrix} S_L \cr C_L \end{pmatrix}  ~~, \cr
\noalign{\kern 10pt}
&&\hskip -1.cm
 S_L  (z; \lambda, -c)  = - S_R  (z; \lambda, c) ~~, \cr
\noalign{\kern 10pt}
&&\hskip -1.cm
C_L C_R - S_L S_R = 1 ~~. 
\label{Bessel7}
\eeqn

\subsection{Equations and the spectrum}
To find the spectrum resulting in the theory with (\ref{Lag4}) and (\ref{Lag5}),  
we start with writing  full equations.  Recall that the Wilson line phase $\theta_H$
mixes $(B, t)$ with $t'$ in the $Q_\EM = \frac{2}{3}$ sector, and $(D, X)$ with $b'$
in the $Q_\EM = -\frac{1}{3}$ sector, respectively.  
The brane mass interactions connect
$B$ to $\hat B_R$, $U$ and $t$ to $\hat U_R$,  in the $Q_\EM = \frac{2}{3}$ sector,
whereas they connect $D$ and $b$ to $\hat D_R$, and $X$ to $\hat X_R$, 
in the $Q_\EM = -\frac{1}{3}$ sector.

Strategy for solving equations in the presence of $\theta_H \not= 0$ is to first
move to a new twisted gauge in which the background field vanishes, 
 ${\tilde A_z}^c = 0$, as described in refs.\ \cite{HS2} and \cite{Falkowski1}.   
 This is achieved by 
\beqn
&&\hskip -1cm
\Omega(z) = \exp \big\{  i \theta (z) \sqrt{2} \, T^{\hat 4} \big\} ~~, \cr
\noalign{\kern 10pt}
&&\hskip -1cm
\theta (z) = \frac{z_L^2 - z^2}{z_L^2 - 1} \, \theta_H ~~, \cr
\noalign{\kern 10pt}
&&\hskip -1cm
g A_z^c = \dot \theta \, \sqrt{2} \, T^{\hat 4} ~~,~~
  \dot \theta = \frac{d\theta}{dz} ~~, \cr
\noalign{\kern 10pt}
&&\hskip -1cm
\tilde \Psi = \Omega (z) \Psi ~~.
\label{GT1}
\eeqn
Note that $\theta(1) = \theta_H$ and $\Omega(z_L) = 1$. 
In the standard vectorial representation $\Psi = (\psi_1, \cdots, \psi_5)^t$,
$\Omega (z)$ takes the form 
\beeq
\Omega(z) = 
\begin{pmatrix} 1 \cr &1 \cr &&1 \cr &&& 
c & s \cr &&& - s & c \end{pmatrix} ~~~,~~~
\begin{cases} c = \cos \theta (z) ~,  \cr s = \sin \theta (z)~. \end{cases}
\label{GT2}
\eneq
In the twisted gauge the equations in the bulk are the same as in free theory,
whereas the boundary conditions at the Planck brane at $z=1$ become
more involved.  In the basis $(B,t,t')$ employed in (\ref{def1}) and (\ref{fermion1})
\beeq
\begin{pmatrix} \tilde B \cr \tilde t \cr \tilde t' \end{pmatrix}
= \tilde \Omega  \begin{pmatrix} B \cr  t \cr  t' \end{pmatrix}  ~~,~~
\tilde \Omega  = 
\begin{pmatrix} 
\onehalf (1+c) & \onehalf (1-c) & - \frac{1}{\sqrt{2}} s\cr  
\onehalf (1-c) & \onehalf (1+c) &  \frac{1}{\sqrt{2}} s\cr  
\frac{1}{\sqrt{2}} s & - \frac{1}{\sqrt{2}} s & c
\end{pmatrix} ~.
\label{GT3}
\eneq
A similar relation applies to $(D, X, b')$. 

\vskip 5pt
\noindent
(i) $Q_\EM = \frac{5}{3}$ sector

In the two component basis 
there are $T_L$, $T_R$ and $\hat T_R$ in this sector.  
As there is no coupling to $\theta_H$,  $\tilde T = T$.
The parity assignments of the bulk fields are $T_L (+,+)$, $T_R (-,-)$. 
With (\ref{Lag4}) and (\ref{Lag5}) equations of motion are given by
\beqn
&&\hskip -1cm
\sigma \dd \, T_L - k D_-(c_1) \, T_R - \mu_1 \delta (y) \,  \hat T_R = 0 ~~, \cr
\noalign{\kern 10pt}
&&\hskip -1cm
\bar \sigma \dd \, T_R -  k D_+(c_1) \, T_L = 0 ~~, \cr
\noalign{\kern 10pt}
&&\hskip -1cm
\bar \sigma \dd \, \hat T_R  - \mu_1 T_L = 0  ~~.
\label{Feq4}
\eeqn
Recall that $D_\pm = (e^{-\sigma}/k) ( \pm d/dy + c d\sigma/dy)$ 
in the $y$ coordinate.  
Integrating the first equation from $y = -\ep$ to $y=+\ep$
and making use of $T_R (x, -y) = - T_R (x, y)$, one finds
\beeq
T_R |_{y= \ep}  = - T_R  |_{y= -\ep} = \frac{\mu_1}{2} \, \hat T_R (x) ~~,
\label{discon1}
\eneq
that is, parity-odd $T_R$ develops a discontinuity at the Planck brane.
Inserting (\ref{discon1}) into the second equation in (\ref{Feq4}) 
and making use of the third equation in (\ref{Feq4}), one finds
$(k D_+(c_1) - \mu_1^2) T_L = 0$ at $y= \ep$.  The boudary conditions
for the bulk field $T_L$ are thus given by
\beeq
\begin{cases}
\bigg( D_+(c_1)  - \myfrac{\mu_1^2}{2 k} \bigg) \,  T_L  = 0 
      &\hbox{at}~ z=1 ~, \cr
\noalign{\kern 10pt}
~ D_+(c_1) \,  T_L  = 0 & \hbox{at}~ z=z_L~. 
\end{cases}
\label{BC3}
\eneq
Boundary conditions for $T_R$ are given by (\ref{discon1}) and 
$T_R |_{z= z_L} =0$, which follow from (\ref{BC3}) and (\ref{Feq4}).  

In the bulk region $1 < z < z_L$, $T_L$ and $T_R$ satisfy free equations.  
Mode functions are obtained with an ansatz 
$T_{L,R} = e^{ipx} u_{L,R} (p) f_{L,R}(z)$ for each mass eigenstate.  
$f_{L,R}(z)$ satisfy $D_+ f_L = \lambda f_R$ and $D_- f_R = \lambda f_L$
so that $T_L$ and $T_R$ satisfy  (\ref{Feq3}).  
The boundary condition (\ref{BC3}) at $z=z_L$
implies that $f_L(z) \propto C_L (z; \lambda , c_1)$.
The boundary condition at $z=1$ is then satisfied if
\beeq
\lambda S_R^{(1)} - \frac{\mu_1^2}{2k} ~  C_L^{(1)}  = 0 ~.
\label{Fspec1}
\eneq
where $S_R^{(j)} = S_R(1; \lambda, c_j) $ etc..
If there were no boundary interaction ($\mu_1=0$),  then the spectrum
contains a zero mode ($\lambda_0 = 0$).    For $\mu_1^2 / 2k \gg \lambda$
the second term dominates over the first term.
The lowest mass $m_0 = k \lambda_0$ determined by 
$C_L(1; \lambda_0, c)  = 0$ is at the Kaluza-Klein mass scale for $c>0$.  
In other words, as long as $\mu_1^2 \gg  m_\KK$,  the mass of the 
lowest mode is $O( m_\KK)$ for $c_1 > 0$.

Here we have been observing an effective change of boundary conditions.  
The Neumann condition corresponding to $\lambda S_R(1; \lambda, c) =0$
changes to the Dirichlet condition corresponding to $C_L(1; \lambda, c)  = 0$
for low-lying modes in the Kaluza-Klein tower.  We note that for $c < - \onehalf$
the lowest mass value  becomes non-vanishing, but remains small.

\vskip 10pt
\noindent
(ii) $Q_\EM = \frac{2}{3}$ sector

In this sector six fields $U$, $B$, $t$, $t'$, $\hat U_R$ and $\hat B_R$
mix with each other.  In the basis $(B, t, t')$ the Wilson line phase $\theta_H$
gives a background field
\beeq
-ig A_z^c = \frac{\dot \theta}{\sqrt{2}}
\begin{pmatrix} && -1 \cr &&1\cr 1& -1 \end{pmatrix} ~~.
\label{Wilson2}
\eneq
We note that $-ig A_z^c = (d \tilde \Omega^\dagger/dz) \tilde \Omega$.
With (\ref{Wilson2})  equations of motion in the original gauge are given by
\beqn
&&\hskip -1cm
\sigma \dd  
\begin{pmatrix}U_L \cr  B_L \cr  t_L \cr  t_L' \end{pmatrix}
- k \begin{pmatrix} D_-^{(2)}\cr &D_-^{(1)} && \frac{1}{\sqrt{2}} \dot\theta \cr 
      &&D_-^{(1)} & - \frac{1}{\sqrt{2}} \dot\theta \cr 
      & - \frac{1}{\sqrt{2}} \dot\theta &  \frac{1}{\sqrt{2}} \dot\theta &D_-^{(1)} \end{pmatrix}
\begin{pmatrix} U_R \cr  B_R \cr  t_R \cr  t_R' \end{pmatrix}
- \delta (y) 
\begin{pmatrix}  \mu_2 \hat U_R \cr \mu_1 \hat B_R \cr
  \tilde \mu ~ \hat U_R \cr 0  \end{pmatrix} = 0 ~,  \cr
\noalign{\kern 15pt}
&&\hskip -1cm
\bar \sigma \dd  
\begin{pmatrix} U_R \cr   B_R \cr  t_R \cr  t_R' \end{pmatrix}
- k \begin{pmatrix} D_+^{(2)}\cr &D_+^{(1)} && -  \frac{1}{\sqrt{2}} \dot\theta  \cr 
&&D_+^{(1)}  &  \frac{1}{\sqrt{2}} \dot\theta  \cr
&  \frac{1}{\sqrt{2}} \dot\theta & -  \frac{1}{\sqrt{2}} \dot\theta &D_+^{(1)} \end{pmatrix}
\begin{pmatrix}  U_L \cr  B_L \cr  t_L \cr  t_L' \end{pmatrix}
=0 ~, \cr
\noalign{\kern 15pt}
&&\hskip -1cm
\bar \sigma \dd 
\begin{pmatrix} \hat U_R \cr \hat B_R \end{pmatrix}
- \begin{pmatrix} \mu_2 & 0 & \tilde \mu  \cr
0 &  \mu_1 &0  \end{pmatrix}
\begin{pmatrix}   U_L \cr   B_L \cr t_L \end{pmatrix}
=0 ~. 
\label{Feq5}
\eeqn
Here $D_\pm^{(j)} = D_\pm (c_j)$.  
Recall that $U_L$, $B_L$, $t_L$ and $t'_R$ have parity $(+,+)$
whereas $U_R$, $B_R$, $t_R$ and $t'_L$ have parity $(-,-)$. 
Integrating the first equation above from $y= -\ep$ to $y= +\ep$ and making use of 
the odd nature of $U_R$, $B_R$ and  $t_R$ under parity, one finds
\beqn
&&\hskip -1cm
U_R |_{y= \ep} = \frac{\mu_2}{2} \, \hat U_R ~~,  \cr
\noalign{\kern 10pt}
&&\hskip -1cm
B_R |_{y= \ep} = \frac{\mu_1}{2} \, \hat B_R ~~,  \cr
\noalign{\kern 10pt}
&&\hskip -1cm
t_R |_{y= \ep} = \frac{\tilde \mu}{2} \, \hat U_R ~~.
\label{discon2}
\eeqn
Another parity-odd field $t'_L$ satisfies $t'_L |_{y=\ep} = 0$ 
as it follows from the second equation in (\ref{Feq5}).  
$U_R$, $B_R$ and  $t_R$ develop discontinuities at the Planck brane,
but $t'_L$ does not.

Now to find boundary conditions for $U_L, B_L,  t_L$ at $y=0$ ($z=1$)
we insert (\ref{discon2}) into the second equation in (\ref{Feq5}) and
use the third equation in (\ref{Feq5}).   Noting that $t'_L$ vanishes there,
one obtains
\beqn
&& \hskip -1cm 
D_+^{(2)} U_L - \frac{\mu_2}{2k} ( \mu_2 U_L + \tilde \mu t_L) = 0 ~, \cr
\noalign{\kern 10pt}
&& \hskip -1cm 
D_+^{(1)} B_L - \frac{\mu_1}{2k} \cdot \mu_1 B_L = 0 ~, \cr
\noalign{\kern 10pt}
&& \hskip -1cm 
D_+^{(1)} t_L - \frac{\tilde \mu}{2k} ( \mu_2 U_L + \tilde \mu t_L) = 0 ~, \cr
\noalign{\kern 10pt}
&& \hskip -1cm 
t'_L = 0 ~~,
\label{BC4}
\eeqn
at $z=1$, and
\beeq
D_+^{(2)} U_L = D_+^{(1)} B_L  = D_+^{(1)} t_L = t'_L = 0 ~~.
\label{BC5}
\eneq
at $z= z_L$. 

At this stage we move to the twisted gauge defined in (\ref{GT1}) and (\ref{GT3})
in which the bulk fields satisfy free equations.  
Taking into account the fact that $\tilde t'_L = t'_L = 0$ at $z=z_L$,
we find that
\beeq
D_+^{(2)} \tilde U_L = D_+^{(1)} \tilde B_L  = D_+^{(1)} \tilde t_L 
= \tilde t'_L = 0 
\label{BC6}
\eneq
at $z= z_L$.  
Making use of (\ref{GT3}) and (\ref{BC4}),  we find
\beqn
&&\hskip -1cm 
s_H (\tilde B_L - \tilde t_L) - \sqrt{2} c_H \tilde t'_L = 0 ~~, \cr
\noalign{\kern 10pt}
&&\hskip -1cm 
\Big( D_+^{(2)} - \frac{\mu_2^2}{2k} \Big) \tilde U_L 
- \frac{\tilde \mu \mu_2}{4k} (\tilde B_L + \tilde t_L)
+  \frac{\tilde \mu \mu_2}{4 c_H k} (\tilde B_L - \tilde t_L) = 0 ~,  \cr
\noalign{\kern 10pt}
&&\hskip -1cm 
- \frac{\tilde \mu \mu_2}{2k} \tilde U_L
+ \Big( D_+^{(1)} - \frac{\mu_1^2 + \tilde \mu^2}{4k} \Big)  (\tilde B_L + \tilde t_L)
- \frac{\mu_1^2 - \tilde \mu^2}{4 c_H k}  (\tilde B_L - \tilde t_L) = 0 ~,  \cr
\noalign{\kern 10pt}
&&\hskip -1cm 
\frac{\tilde \mu \mu_2}{2k} \tilde U_L
- \frac{\mu_1^2 - \tilde \mu^2}{4k}  (\tilde B_L + \tilde t_L)
+ \Big( c_H D_+^{(1)} -  \frac{\mu_1^2 + \tilde \mu^2}{4 c_H k} \Big) 
 (\tilde B_L - \tilde t_L) \cr
\noalign{\kern 10pt}
&&\hskip 6.5cm 
+ \sqrt{2} s_H D_+^{(1)} \tilde t'_L = 0 
\label{BC7}
\eeqn
at $z=1$ where  $c_H = \cos \theta_H$ and  $s_H = \sin \theta_H$.  
All the fields satisfy free equations.  With the boundary conditions (\ref{BC6})
at $z= z_L$, mode functions can be expressed as
\beqn
&&\hskip -1cm
\tilde U_L = a_U \, C_L(z; \lambda, c_2) ~,  \cr
\noalign{\kern 5pt}
&&\hskip -1cm
\tilde B_L \pm \tilde t_L = a _{B \pm t} \, C_L(z; \lambda, c_1) ~, \cr
\noalign{\kern 5pt}
&&\hskip -1cm
\tilde t'_L  = a _{t'} \, S_L(z; \lambda, c_1) ~.
\label{mode1}
\eeqn
Eigenvalues for $\lambda$ are determined by the boundary conditions
(\ref{BC7}) at $z=1$.    Inserting (\ref{mode1}) into (\ref{BC7}), one finds
after lengthy but straightforward manipulation that
\beqn
&&\hskip -1cm
\lambda  A_1 (\lambda) + \lambda B_1 (\lambda) \sin^2 \theta_H   = 0 ~, \cr
\noalign{\kern 10pt}
&&\hskip -1cm
A_1 (\lambda) = \Big( \lambda S_R^{(1)} - \frac{\mu_1^2}{2k} C_L^{(1)} \Big) 
\Big( \lambda S_R^{(1)} S_R^{(2)} 
 -\frac{\tilde \mu^2}{2k} C_L^{(1)} S_R^{(2)}  
 -\frac{\mu_2^2}{2k}S_R^{(1)}  C_L^{(2)}  \Big) ~, \cr
\noalign{\kern 10pt}
&&\hskip -1cm
B_1 (\lambda) = \frac{1}{ S_L^{(1)}} \Big( \lambda^2 S_R^{(1)} S_R^{(2)}
- \lambda \frac{\mu_2^2}{2k} S_R^{(1)}  C_L^{(2)}
- \lambda \frac{\mu_1^2 + \tilde\mu^2}{4k}  C_L^{(1)} S_R^{(2)} 
+  \frac{\mu_1^2 \mu_2^2}{8 k^2}  C_L^{(1)} C_L^{(2)} \Big) 
\label{Fspec2}
\eeqn
where $S_R^{(j)} = S_R(1; \lambda, c_j)$ etc..  

When boundary masses $\mu_j$ and $\tilde \mu$ vanish, (\ref{Fspec2}) reduces
to
\beeq
\lambda S_R^{(1)} \cdot \lambda \Big( S_R^{(1)} + \frac{\sin^2 \theta_H}{S_L^{(1)}} \Big)
\cdot \lambda S_R^{(2)}  = 0 ~.
\label{Fspec3}
\eneq
The first and third factors correspond to  the KK towers of $t+B$ and $U$, respectively.  
The second factor, yielding $S_L^{(1)} S_R^{(1)} + \sin^2 \theta_H =0$ for
$\theta_H \not= 0$,  gives a spectrum of the KK towers of $t-B$ and $t'$.  The zero mode 
of $t-B$ acquires a mass by $\theta_H \not= 0$, but the zero modes of $t+B$ and $U$
remain massless.   With nonvanishing boundary masses these unwanted light modes
acquire masses of $O(m_\KK)$ for $c_1, c_2 >0$.


\vskip 10pt
\noindent
(iii) $Q_\EM = - \frac{1}{3}$ sector

This sector has a similar structure to that in the $Q_\EM =  \frac{2}{3}$ sector.
Six fields $b, D, X, b', \hat D_R$ and $\hat X_R$ mix with each other. 
$b_L$, $D_L$, $X_L$ and $b'_R$ have parity $(+,+)$
whereas $b_R$, $D_R$, $X_R$ and $b'_L$ have parity $(-,-)$. 
Equations of motion are given by
\beqn
&&\hskip -1cm
\sigma \dd  
\begin{pmatrix}b_L \cr  D_L \cr  X_L \cr  b_L' \end{pmatrix}
- k \begin{pmatrix} D_-^{(1)}\cr &D_-^{(2)} && \frac{1}{\sqrt{2}} \dot\theta \cr 
      &&D_-^{(2)} & - \frac{1}{\sqrt{2}} \dot\theta \cr 
      & - \frac{1}{\sqrt{2}} \dot\theta &  \frac{1}{\sqrt{2}} \dot\theta &D_-^{(2)} \end{pmatrix}
\begin{pmatrix} b_R \cr  D_R \cr  X_R \cr  b_R' \end{pmatrix}
- \delta (y) 
\begin{pmatrix}  \tilde \mu ~ \hat D_R \cr \mu_2 \hat D_R \cr
   \mu_3   \hat X_R \cr 0  \end{pmatrix} = 0 ~,  \cr
\noalign{\kern 15pt}
&&\hskip -1cm
\bar \sigma \dd  
\begin{pmatrix} b_R \cr   D_R \cr  X_R \cr  b_R' \end{pmatrix}
- k \begin{pmatrix} D_+^{(1)}\cr &D_+^{(2)} && -  \frac{1}{\sqrt{2}} \dot\theta  \cr 
&&D_+^{(2)}  &  \frac{1}{\sqrt{2}} \dot\theta  \cr
&  \frac{1}{\sqrt{2}} \dot\theta & -  \frac{1}{\sqrt{2}} \dot\theta &D_+^{(2)} \end{pmatrix}
\begin{pmatrix}  b_L \cr  D_L \cr  X_L \cr  b_L' \end{pmatrix}
=0 ~, \cr
\noalign{\kern 15pt}
&&\hskip -1cm
\bar \sigma \dd 
\begin{pmatrix} \hat D_R \cr \hat X_R \end{pmatrix}
- \begin{pmatrix} \tilde \mu  & \mu_2 & 0  \cr
0 &  0 & \mu_3  \end{pmatrix}
\begin{pmatrix}   b_L \cr   D_L \cr X_L \end{pmatrix}
=0 ~. 
\label{Feq6}
\eeqn
This time $b_R, D_R$ and $X_R$ develop discontinuities at the Planck brane:
\beqn
&&\hskip -1cm
b_R |_{y= \ep} = \frac{\tilde \mu}{2} \, \hat D_R ~~,  \cr
\noalign{\kern 10pt}
&&\hskip -1cm
D_R |_{y= \ep} = \frac{\mu_2}{2} \, \hat D_R ~~,  \cr
\noalign{\kern 10pt}
&&\hskip -1cm
X_R |_{y= \ep} = \frac{\mu_3}{2} \, \hat X_R ~~.
\label{discon3}
\eeqn
With (\ref{discon3}) boundary conditions for the left-handed bulk fields are 
found to be
\beqn
&& \hskip -1cm 
D_+^{(1)} b_L - \frac{\tilde \mu}{2k} \, ( \tilde \mu \, b_L + \mu_2 D_L) = 0 ~, \cr
\noalign{\kern 10pt}
&& \hskip -1cm 
D_+^{(2)} D_L - \frac{\mu_2}{2k} \, ( \tilde \mu \,  b_L + \mu_2 D_L)  = 0 ~, \cr
\noalign{\kern 10pt}
&& \hskip -1cm 
D_+^{(2)} X_L - \frac{\mu_3}{2k} \cdot  \mu_3 X_L  = 0 ~, \cr
\noalign{\kern 10pt}
&& \hskip -1cm 
b'_L = 0 ~~,
\label{BC8}
\eeqn
at $z=1$, and $D_+^{(1)} b_L = D_+^{(2)} D_L  = D_+^{(2)} X_L = b'_L = 0$
at $z= z_L$. 
In the twisted gauge,  mode functions are expressed,  as in (\ref{mode1}), as
\beqn
&&\hskip -1cm
\tilde b_L = a_b \, C_L(z; \lambda, c_1) ~,  \cr
\noalign{\kern 5pt}
&&\hskip -1cm
\tilde D_L \pm \tilde X_L = a _{D \pm X} \, C_L(z; \lambda, c_2) ~, \cr
\noalign{\kern 5pt}
&&\hskip -1cm
\tilde b'_L  = a _{b'} \, S_L(z; \lambda, c_2) ~.
\label{mode2}
\eeqn
The boundary conditions at $z=1$, (\ref{BC8}), are satisfied if 
\beqn
&&\hskip -1cm
\lambda  A_2 (\lambda) + \lambda B_2 (\lambda) \sin^2 \theta_H   = 0 ~, \cr
\noalign{\kern 10pt}
&&\hskip -1cm
A_2 (\lambda) = \Big( \lambda S_R^{(2)} - \frac{\mu_3^2}{2k} C_L^{(2)} \Big) 
\Big( \lambda S_R^{(1)} S_R^{(2)} 
 -\frac{\tilde \mu^2}{2k} C_L^{(1)} S_R^{(2)}  
 -\frac{\mu_2^2}{2k}S_R^{(1)}  C_L^{(2)}  \Big) ~, \cr
\noalign{\kern 10pt}
&&\hskip -1cm
B_2 (\lambda) = \frac{1}{ S_L^{(2)}} \Big( \lambda^2 S_R^{(1)} S_R^{(2)}
- \lambda  \frac{\mu_2^2 + \mu_3^2}{4k}  S_R^{(1)}  C_L^{(2)}
- \lambda \frac{\tilde\mu^2}{2k}  C_L^{(1)} S_R^{(2)} 
+  \frac{\tilde \mu^2 \mu_3^2}{8 k^2}  C_L^{(1)} C_L^{(2)} \Big) ~.
\label{Fspec4}
\eeqn

There is a subtle difference between the $Q_\EM = \frac{2}{3}$ and $-\frac{1}{3}$
sectors.
The expression (\ref{Fspec4}) can be obtained from  (\ref{Fspec2})  by formally
interchanging $(c_1, c_2)$,  $(\mu_1, \mu_3)$, and $(\mu_2, \tilde \mu)$.  
$\hat \chi_{2R}$, which was introduced to lift the lowest mode of 
$Q_{2L}=(U_L, D_L)$  of $\Psi_2$ to the KK scale with the mass term $\mu_2$,  
also couples to $q_L = (t_L, b_L)$ of $\Psi_1$ with the mass term  $\tilde \mu$.
It is, therefore,  natural to suppose that $\tilde \mu^2 \ll \mu_j^2$.  
We will see below that this  leads to $m_b \ll m_t$ as desired.

\vskip 10pt
\noindent
(iv) $Q_\EM = - \frac{4}{3}$ sector

$Y$ and $\hat Y_R$ belong to this sector.  Equations of motion are
obtained from (\ref{Feq4}) by replacing $(T, \hat T_R)$ by $(Y, \hat Y_R)$,
and $(c_1, \mu_1)$ by $(c_2, \mu_3)$.  The spectrum is determined by
\beeq
\lambda S_R^{(2)} - \frac{\mu_3^2}{2k} C_L^{(2)} = 0 ~~.
\label{Fspec5}
\eneq


\subsection{Top and bottom masses}

The fermion mass spectrum is determined by the relations 
(\ref{Fspec1}),  (\ref{Fspec2}), 
(\ref{Fspec4}) and (\ref{Fspec5}).  The brane mass terms are expected to 
emerge when the RS warped spacetime is generated at high energy scale.
Even though we do not know how they emerge, it is natural to imagine that
all $\mu_j^2$ and $\tilde \mu^2$ are at that high scale, namely of $O(M_\GUT)$
or $O(M_\Planck)$.  What we need and assume in the present paper is 
much more modest.  We only suppose  that $\mu_j^2, \tilde \mu^2 \gg m_\KK$.  
It will be seen below that the only relevant parameter for low energy physics is 
$\tilde\mu / \mu_2$ in this case.  

For low-lying modes in the Kaluza-Klein towers $m= \lambda k \ll \mu_j^2$ so that
(\ref{Fspec1}) and (\ref{Fspec5}) in the $Q_\EM = \frac{5}{3}, - \frac{4}{3}$ sectors
are approximated by $C_L^{(1)}=0$ and $C_L^{(2)}=0$, respectively.  
The lowest mode  in each sector has a mass of $O(m_\KK)$ for $c_j >0$.  

In a similar manner the relation  (\ref{Fspec2}) in the $Q_\EM = \frac{2}{3}$ sector 
is approximated by
\beqn
&&\hskip -1cm
C_L^{(1)}=0 ~, \cr
\noalign{\kern 10pt}
&&\hskip -1cm
\mu_2^2 C_L^{(2)} \bigg\{ S_R^{(1)} + \frac{\sin^2 \theta_H}{2 S_L^{(1)}} \bigg\} 
+ \tilde \mu^2 C_L^{(1)}S_R^{(2)} =0 ~.
\label{Fspec6}
\eeqn
The first one gives a KK tower of $(B, \hat B_R)$.  The second one contains 
towers of $(t, t')$ and $(U, \hat U_R)$.  It is found below that $m_t$ and $m_b$
can be reproduced  if $c_1 \sim c_2$.  In the limiting case $c_1=c_2$ the second one
splits into $C_L^{(1)} = 0$ for $(U, \hat U_R)$ and 
\beeq
2 \Big( 1 + \frac{\tilde \mu^2}{\mu_2^2} \Big) S_L^{(1)} S_R^{(1)} + \sin^2 \theta_H = 0 
\label{Fspec7}
\eneq
for $(t, t')$.  Notice the appearance of a factor 2 in (\ref{Fspec7}) compared with 
the similar expression in the middle of (\ref{Fspec3}) due to the brane interactions.

The spectrum determined by (\ref{Fspec6}) contains one light mode, namely a top 
quark.  For $\lambda z_L \ll 1$ and $0 < c < \onehalf$, 
$C_L(1; \lambda, c) \sim  z_L^c$ and
$S_L(1; \lambda, c) \sim  -\lambda z_L^{1+ c}/(1+ 2c)$.
Making use of these relations, one finds the top quark mass, $m_t = \lambda k$, 
to be given,  for $0< c_1, c_2 < \onehalf$, by
\beqn
m_t &\sim& \frac{m_\KK}{\sqrt{2}\pi}
\frac{\sqrt{1 - 4c_1^2} \, |\sin\theta_H|} 
{\bigg( 1  +  \myfrac{\tilde \mu^2}{\mu_2^2} 
     \myfrac{1-2c_1}{1-2c_2}  \, z_L^{2(c_1 -c_2)} \bigg)^{1/2}} \cr
\noalign{\kern 10pt}
&\sim&  \frac{m_\KK}{\sqrt{2}\pi} \, \sqrt{1 - 4c_1^2} ~  |\sin\theta_H|
\quad \hbox{for } \frac{\tilde\mu^2}{\mu_2^2} \,  z_L^{2(c_1-c_2)} \ll 1~.
\label{top1}
\eeqn
With $\theta_H = \pm \onehalf \pi$, we find $c_1 \sim 0.43$ for $m_t = 172 \,$GeV.

The spectrum  in the $Q_\EM = - \frac{1}{3}$ sector is obtained in a similar manner.  
(\ref{Fspec5}) is approximated by
\beqn
&&\hskip -1cm
C_L^{(2)}=0 ~, \cr
\noalign{\kern 10pt}
&&\hskip -1cm
\tilde \mu^2 C_L^{(1)} \bigg\{ S_R^{(2)} + \frac{\sin^2 \theta_H}{2 S_L^{(2)}} \bigg\} 
+ \mu_2^2 C_L^{(2)}S_R^{(1)} =0 ~.
\label{Fspec8}
\eeqn
There exists one light mode, which is identified with  the bottom quark.  
Its mass is given,  for $0< c_1, c_2 < \onehalf$,  by
\beqn
m_b &\sim& \frac{m_\KK}{\sqrt{2}\pi}
\frac{\sqrt{1 - 4c_2^2} \, |\sin\theta_H|} 
{\bigg( 1  + \myfrac{ \mu_2^2}{\tilde \mu^2} 
     \myfrac{1-2c_2}{1-2c_1}  \, z_L^{2(c_2 -c_1)} \bigg)^{1/2}}   \cr
\noalign{\kern 10pt}
&=& \sqrt{\frac{1+2c_2}{1+2c_1}} ~ \Big| \frac{\tilde \mu}{\mu_2} \Big| \,
z_L^{c_1 - c_2} \cdot m_t
\label{bottom1}
\eeqn
which justifies the approximation employed in (\ref{top1}).
$|c_1 - c_2|$ must be small to get a reasonable value for $\tilde\mu/\mu_2$.
In the most attractive scenario  $c_1 = c_2$, which results if all $t$, $t'$, $b$, and $b'$ 
belong to one single multiplet in a larger unified theory,  one finds that
\beeq
\Big| \frac{\tilde\mu}{\mu_2} \Big| = \frac{m_b}{m_t} \ll 1   ~.
\label{topbottom1}
\eneq
We stress that only the ratio $\tilde\mu/\mu_2$ among the brane masses
is relevant for  $m_b$.
Individual values of the brane mass parameters $\mu_j^2, \tilde \mu^2$ are 
irrelevant so long as they are much bigger than $m_\KK$.
To have non-vanishing $m_b$ we need both $\theta_H \not= 0$ and 
$\tilde\mu \not= 0$.
$b_L$ in $\Psi_1$ must be connected with $b'_R$ in $\Psi_2$.

One may wonder if there are other values for $c_1$ and $c_2$ to reproduce $m_t$ 
and $m_b$.  In the cases $0< c_1 < \onehalf < c_2$ and 
$-\onehalf < c_2 < 0< c_1 < \onehalf$,
one obtains the same relation for 
$m_b/m_t$ as the second relation in (\ref{bottom1}), which demands unnaturally large
$\tilde\mu/\mu_2$ as $z_L \sim 10^{15}$.
In the current scheme, the observed $m_t$ and $m_b$ are realized only for 
$0 < c_1, c_2 < \onehalf$.  

It is straightforward to incorporate light quarks in the first and second generations.
For each generation two {\bf 5} multiplets and associated brane fermions
are introduced.  The bulk mass parameter $c_1=c_2 \equiv c$ and  
brane masses $\mu_1, \mu_2, \mu_3$ and $\tilde \mu$
take  values depending on the generation.
As their masses are much smaller than $m_W$, it will be found that $c > \onehalf$.
The spectrum is determined by the equations (\ref{Fspec6}) and (\ref{Fspec8}).
For up- and down-type quarks we find, for  $c > \onehalf$, that
\beqn
m_{\rm up} 
&\sim& \frac{m_\KK}{\sqrt{2}\pi}
\frac{\sqrt{4c^2 -1} \, |\sin\theta_H|} 
{\bigg( 1 + \myfrac{\tilde \mu^2}{\mu_2^2} \bigg)^{1/2} ~
z_L^{c - \onehalf} }     \cr
\noalign{\kern 10pt}
m_{\rm down}
&\sim&  \bigg| \frac{\tilde \mu}{\mu_2} \bigg| ~ m_{\rm up} ~~.
\label{updownmass}
\eeqn
With $\theta_H = \pm \onehalf \pi$ and $z_L=10^{15}$, 
we find $c \sim 0.653$ and $0.853$ for
$m_c = 1.4\,$GeV and $m_u = 4\,$MeV, respectively.
A similar construction is done for leptons by putting $(e^c_R, \nu^c_R, e^c_L)$
and $\nu^c_L$ in $\Psi_1$ and $\Psi_2$, respectively.
Large hierarchy in fermion masses can be naturally explained by modest
distribution in the bulk mass parameter $c$, as was  pointed out 
in ref.\  \cite{Arkani} in general context and in  ref.\ \cite{HNSS} in the 
gauge-Higgs unification scenario.

\section{Dynamical EW symmetry breaking}

The value for $\theta_H$ is determined by the location of the global
minimum of the effective potential $V_\eff (\theta_H)$, which becomes
nontrivial at the quantum level.\cite{YH1}  When $\theta_H$ takes a 
nontrivial value, the standard model symmetry $SU(2)_L \times U(1)_Y$
dynamically breaks down to $U(1)_\EM$.  In pure gauge theory without fermions
the symmetry remains unbroken.  We shall show below that the presence of
a top quark induces the symmetry breaking.

The evaluation of the effective potential $V_\eff (\theta_H)$ in the RS warped 
spacetime was initiated by Oda and Weiler.\cite{Oda1}  
Since then a powerful method for
the evaluation has been developed  by Falkowski.\cite{Falkowski1} 
Concrete evaluation in the gauge-Higgs unification models of electroweak interactions 
in the RS spacetime has been given in refs.\ \cite{Wagner1} and \cite{Hatanaka1}.

The effective potential $V_\eff (\theta_H)$ at the one loop level is determined 
by the dependence of the mass spectrum on $\theta_H$.  We have seen in the 
preceding sections that spectra  in both gauge-Higgs and
fermion sectors are determined by zeros of
equations of the type $A(\lambda) + B(\lambda) f(\theta_H) =0$.  
For gauge fields and fermions in the vector representation, we have seen 
$f(\theta_H) = \sin^2 \theta_H (\equiv f_1(\theta_H))$.   
For matter fields in the spinor representation
one finds $f(\theta_H) = \sin^2 \onehalf \theta_H$.  (See, for example, 
refs.\ \cite{SH1} and  \cite{HS2}.)

We rewrite the equation in the form $1 + \tilde Q(\lambda)  f(\theta_H) =0$
($\tilde Q = B/A$), which yields a spectrum $\{ \lambda_n (\theta_H)  \}$.  
Then one-loop contribution to $V_\eff (\theta_H)$ coming from particles
with masses $m_n(\theta_H)  = k \lambda_n$  is given by
\beqn
&&\hskip -1cm
V_\eff (\theta_H) = \pm \frac{1}{2} \int \frac{d^4 p}{(2\pi)^4} 
\sum \ln \big( p^2 + m_n(\theta_H)^2  \big) \cr
\noalign{\kern 10pt}
&&\hskip 0.5cm
= \pm  I [ Q(q) ; f(\theta_H) ] \cr
\noalign{\kern 10pt}
&&\hskip -1cm
 I [ Q(q) ; f(\theta_H) ]=  \frac{(k z_L^{-1})^4}{(4\pi)^2} \int_0^\infty dq  \, q^3 
\ln \Big\{ 1 + Q(q) f(\theta_H) \Big\} ~,  \cr
\noalign{\kern 10pt}
&&\hskip - 1.cm
Q(q) = \tilde Q (i q z_L^{-1}) ~.
\label{effV1}
\eeqn
Here $\pm$ corresponds to bosons or fermions, and $\theta_H$-independent
constant terms have been ignored.  It will be seen below that the integral is
dominated by the integrand in a range $0 < q < 10$.

\subsection{Contributions from the gauge field sector}

\begin{figure}[t,b]
\centering  \leavevmode
\includegraphics[height=5cm]{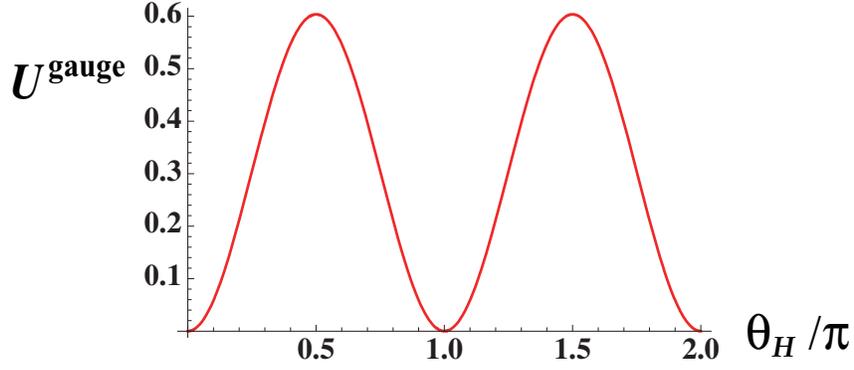}
\caption{The effective potential  $V_\eff (\theta_H)^{\rm gauge}$ in pure gauge theroy without fermions. The plot is for  
$U^{\rm gauge}(\theta_H/\pi) = (4\pi)^2 (kz_L^{-1})^{-4} \, V_\eff^{\rm gauge}$ 
at  $z_L = 10^{15}$.}
\label{effV-gauge-fig}
\end{figure}

The $W$ tower, (\ref{gHspec1b}),  the $Z$ tower, (\ref{gHspec2c}), and
the scalar tower, (\ref{scalarspec2}),  with associated ghost contributions, 
contribute to $V_\eff (\theta_H)$.  
Let us define 
\beqn
&&\hskip -1cm
\hat F_{\alpha ,\beta} (u, v) = I_\alpha (u) K_\beta(v) 
- e^{- i (\alpha - \beta) \pi}  K_\alpha (u) I_\beta(v)  ~~, \cr
\noalign{\kern 10pt}
&&\hskip -1cm
F_{\alpha , \beta} (iu, iv) 
= - \frac{2}{\pi} e^{i (\alpha - \beta) \pi/2} \hat  F_{\alpha ,\beta} (u, v)  
\label{Bessel8}
\eeqn
where $I_\alpha$ and $K_\alpha$ are modified Bessel functions and
$F_{\alpha , \beta}$ is defined in (\ref{Bessel3}).  
The effective potential is given by
\beqn
&&\hskip -1cm
V_\eff (\theta_H)^{\rm gauge} = 
2 \cdot 2 \cdot I [ \onehalf Q_0(q, \onehalf) ;  f_1(\theta_H) ] 
+ 2 \cdot I [ \onehalf (1 + s_\phi^2) Q_0(q, \onehalf) ;  f_1(\theta_H) ]  \cr
\noalign{\kern 10pt}
&&\hskip 5cm 
+ 3 \cdot I [ Q_0(q, \onehalf) ;  f_1(\theta_H) ] ~~, \cr
\noalign{\kern 10pt}
&&\hskip -1cm 
Q_0(q, c) = \frac{z_L}{q^2} 
\frac{1}{\hat F_{c - \onehalf,c - \onehalf} (qz_L^{-1}, q) 
              \hat F_{c + \onehalf, c + \onehalf} (qz_L^{-1}, q) }  ~~, \cr
\noalign{\kern 10pt}
&&\hskip -1cm 
f_1(\theta_H) = \sin^2 \theta_H  ~~. 
\label{effV2}
\eeqn
The behavior of $V_\eff (\theta_H)^{\rm gauge}$ is depicted 
in fig.\ \ref{effV-gauge-fig}.
It has global minima at $\theta_H = 0$ and $\pi$.  $SU(2)_L \times U(1)_Y$
symmetry remains unbroken in pure gauge theory.

The flat spacetime limit $k \go 0$ of $V_\eff (\theta_H)^{\rm gauge}$ is
obtained by replacing $k z_L^{-1} \sim m_\KK/\pi$  by $1/L$, and 
$Q_0(q, \onehalf)$ by $Q_{\rm flat}(q) = 1/ \sinh^2 q$.  
The shape of $V_\eff^{\rm gauge}$  in the RS space is similar to that 
in flat space.  The magnitude of 
$U^{\rm gauge} = (4\pi)^2 (kz_L^{-1})^{-4} \, V_\eff^{\rm gauge}$ 
in RS is reduced compared 
to that in flat space by a factor $2/kL$.

\subsection{Contributions from the fermion sector}

In the fermion sector the spectra in the $Q=\frac{2}{3}$ and $- \frac{1}{3}$ sectors,  
(\ref{Fspec2}) and (\ref{Fspec4}),  yield non-trivial contributions
to $V_\eff (\theta_H)$.  $\tilde Q$ is given by $B_1/A_1$ or $B_2/A_2$
in each sector.  When there were no boundary masses, $\mu_j , \tilde \mu = 0$,
then $\tilde Q$ would take the form $1/ S_L^{(j)} S_R^{(j)} $ ($j=1,2$).  It immediately
follows that
\beqn
V_\eff (\theta_H)^{\rm fermion} \Big|_{\mu_j, \tilde\mu=0}
= - 4 \, \Big\{  I [ Q_0(q, c_1); f_1(\theta_H) ] + 
   I [ Q_0(q, c_2); f_1(\theta_H) ] \Big\} ~~.
\label{effV3}
\eeqn
Here  the  factor 4 accounts for the number of degrees of freedom. 
We have seen that $c_1 \sim 0.43$ for the top quark multiplet.   
With this value the magnitude of the contribution from the top quark 
multiplet ($-4 I[Q_0; f_1]$) is three times larger than that of 
$V_\eff (\theta_H)^{\rm gauge}$  in (\ref{effV2}).  
The global minima are found at $\theta_H = \pm \onehalf \pi$,
which implies the EW symmetry breaking, although with vanishing 
 $\mu_j , \tilde \mu$ there appear unwanted massless particles.
We remark that
a contribution $ I [ Q_0(q, c); f_1(\theta_H) ]$ becomes negligible 
for $c > 0.6$ compared with the gauge field contributions.  
As a consequence contributions from light quarks and leptons 
become negligibly small in the RS space.  

To get   $Q_j (q)$ for $\mu_j, \tilde \mu \not= 0$ 
from $\tilde Q_j (\lambda) = B_j/A_j$,  it is sufficient to make 
replacement 
\beqn
\lambda ~~  &\go& i q z_L^{-1} ~, \cr
\noalign{\kern 10pt}
\begin{pmatrix} S_L \cr S_R \end{pmatrix} 
&\go& 
\pm iq z_L^{-1/2} \, \hat F_{c \pm \onehalf , c \pm \onehalf} (qz_L^{-1}, q) ~, \cr
\noalign{\kern 10pt}
\begin{pmatrix} C_L \cr C_R \end{pmatrix} 
&\go&
q z_L^{-1/2} \, \hat F_{c \pm \onehalf , c \mp \onehalf} (qz_L^{-1}, q) ~.
\label{replace1}
\eeqn
The resultant expressions for $Q_j(q)$'s are not illuminating.
When $\mu_j^2, \tilde \mu^2 \gg m_\KK$,  they tremendously 
simplify.  In particular   for $c_1 = c_2 = c$ they  become
\beqn
&&\hskip -1cm
Q_1 (q) \simeq \frac{\mu_2^2}{2(\mu_2^2 + \tilde \mu^2)} \, Q_0 (q, c) ~, \cr
\noalign{\kern 10pt}
&&\hskip -1cm
Q_2 (q) \simeq \frac{\tilde \mu^2}{2(\mu_2^2 + \tilde \mu^2)} \, Q_0 (q, c) ~,
\label{effV4}
\eeqn
The approximation is valid for $q \ll \mu_j^2/m_\KK, \tilde \mu^2/m_\KK$.
As $Q_j(q)$ becomes negligibly small for $q > 10$, the expression 
(\ref{effV4}) can be safely used in the integral $I[Q_j (q) , f_1 (\theta_H)]$
for numerical evaluation. 
One finds
\beqn
&&\hskip -1cm
V_\eff (\theta_H)^{\rm fermion} \simeq -4 \bigg\{ 
I \Big[ \frac{1}{2(1+r)} Q_0 (q, c) ; f_1(\theta_H) \Big] 
+ I \Big[ \frac{r}{2(1+r)} Q_0 (q, c) ; f_1(\theta_H) \Big] \bigg\} ~, \cr
\noalign{\kern 10pt}
&&\hskip 1cm
r = \frac{\tilde \mu^2}{\mu_2^2} = \Big( \frac{m_b}{m_t} \Big)^2 ~~.
\label{effV5}
\eeqn
As $r \ll 1$, the first term in (\ref{effV5}) coming from $\Psi_1$ dominates.  
The factor $\onehalf$ in the argument of $I$ is due to the fact that 
$t'$ couples through $\theta_H$ to $\psi_4 = (t -B)/\sqrt{2}$, 
and the $B$ component becomes heavy.  As for the $\Psi_2$ contribution, 
both $D$ and $X$, the partners of $b'$, become heavy so that no component
except for a small mixture of $b$ characterized by a factor $r$ remains light.
This accounts for the difference between (\ref{effV3}) and  (\ref{effV5}).  

\subsection{Symmetry breaking}

\begin{figure}[t,b]
\centering  \leavevmode
\includegraphics[height=5cm]{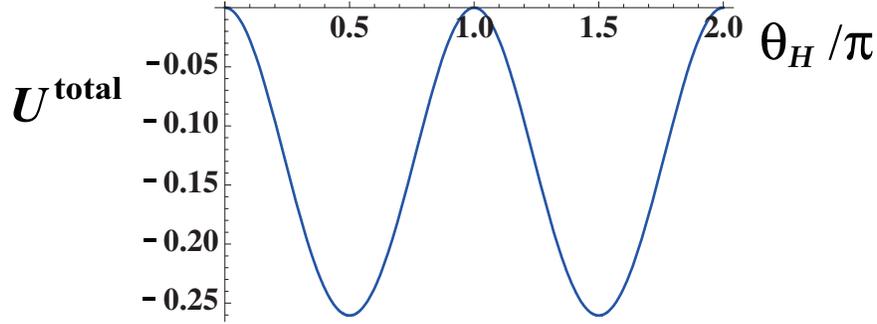}
\caption{The effective potential  $V_\eff (\theta_H)$ in the model  with
top and bottom quarks. The plot is for  
$U^{\rm total} (\theta_H/\pi) = (4\pi)^2 (kz_L^{-1})^{-4} \, V_\eff$ 
at  $z_L = 10^{15}$.  Contributions from light quarks and leptons 
are negligible.
The global minima are located at 
$\theta_H = \onehalf \pi $ and $\frac{3}{2}\pi$, where  the EW symmetry  
dynamically breaks down to $U(1)_{\EM}$.}
\label{effV-total-fig}
\end{figure}

The total effective potential $V_\eff (\theta_H)$  is the sum of 
 $V_\eff (\theta_H)^{\rm gauge}$ in (\ref{effV2}) and 
 $V_\eff (\theta_H)^{\rm fermion}$ in (\ref{effV5}).
 It is displayed in fig.\ \ref{effV-total-fig}.
With $c \sim 0.43$ the top contribution dominates over others.  
$V_\eff (\theta_H)$ has global minima at $\theta_H = \pm \onehalf \pi$, 
where the EW symmetry dynamically breaks down to $U(1)_\EM$.  
The contributions from  other quarks and leptons are
negligible,  as the corresponding bulk mass parameters $c$ range from
0.6 to 0.9.\cite{HNSS}  We conclude that the presence of a heavy top quark 
triggers EW symmetry breaking. 

The effective potential $V_\eff (\theta_H)$ depends on the parameter $z_L$.
There are two critical values for $z_L$.  As $z_L$ is decreased, the value of 
the bulk mass parameter $c$ also decreases to reproduce the observed $m_t$.  
At $z_L = 9.4 \times 10^3 \equiv z_L^{c1}$,
which corresponds to $k = 2.3 \times 10^6 \,$GeV, $c$ becomes 0.
For $z_L < z_L^{c1}$ there exists no solution with the observed $m_t$.
One can set $c$ to be 0 and examine the behavior of $V_\eff (\theta_H)$
for $z_L < z_L^{c1}$.  It is found that for $z_L < z_L^{c2} = 905$, 
the global minima of $V_\eff (\theta_H)$ shift to $\theta_H = 0$ and $\pi$
so that the EW symmetry is unbroken.
One may take the flat space limit ($k \go 0$) with the bulk mass $ck$ kept fixed.
In this case $c \go \infty$ as $k\go 0$ so that contributions of fermions to the 
effective potential are exponentially suppressed.  
We conclude that the EW symmetry is unbroken in flat space in our scheme.

We would also like to remark that if fermions were introduced in the spinor representation
of $SO(5)$, then there would be no EW symmetry breaking.  In the effective
potential fermions would give $f(\theta_H) = \sin^2 \onehalf \theta_H$ 
in the expression (\ref{effV1})   so that
the global minimum would appear either at $\theta_H = 0 $ or $\pi$.

\section{Higgs mass}

The four-dimensional Higgs field (\ref{Higgs1}) acquires a finite mass at the 
one loop level.  The physical neutral Higgs field $\phi^4 \equiv \phi_H$ is related to the 
Wilson line phase $\theta_H$ by (\ref{Wilson1}).  The effective potential 
$V_\eff (\theta_H)$ evaluated in the previous section translates to
the effective potential for the Higgs field $\phi_H$.  By expanding $V_\eff$ around
the minimum one obtains
\beqn
&&\hskip -1cm
V_\eff = \hbox{const.} + \frac{1}{2} m_H^2 (\phi_H - v)^2 + \cdots ~, \cr
\noalign{\kern 10pt}
&&\hskip -1cm
m_H^2 =  \frac{\pi^2 g_4^2 kL}{4 \,  m_\KK^2}  \, 
\frac{d^2 V_\eff}{d \theta_H^2} \bigg|_{\rm min} ~~.
\label{Higgs2}
\eeqn
We recall that
\beeq
v = \la \phi_H \ra = \frac{2 \theta_H}{\pi g_4} \frac{m_\KK}{\sqrt{kL}} 
=\frac{2}{g_4} \frac{\theta_H}{|\sin \theta_H |} \, m_W ~.
\eneq
The relation between $v$ and $m_W$ deviates from that in the standard 
model by a factor $\onehalf \pi$ at the global minimum 
$\theta_H = \pm \onehalf \pi$.  

Inserting (\ref{effV2}) and (\ref{effV5}) into (\ref{Higgs2}), we find
\beqn
&&\hskip -1cm
m_H^2 \simeq \frac{g_4^2 kL m_\KK^2}{64 \pi^4} 
\bigg\{ - 4 G[\onehalf Q_0(q, \onehalf)] -  2 G[\onehalf (1+s_\phi^2)Q_0(q, \onehalf)] 
- 3  G[Q_0(q, \onehalf)]  \cr
\noalign{\kern 10pt}
&&\hskip 3cm
+ 4 G \Big[ \frac{1}{2(1+r)} Q_0(q,c) \Big] 
+ 4 G \Big[ \frac{r}{2(1+r)} Q_0(q,c) \Big] \bigg\} ~~. \cr
\noalign{\kern 10pt}
&&\hskip -1cm
G[Q(q)] = \int_0^\infty dq \, q^3 \frac{2 Q(q)}{1 + Q(q)}  ~~.
\eeqn
The contribution from the bottom quark (the last term in the parenthesis) 
to $m_H^2$ is negligible. 
With numerical values $m_W$, $m_t$, $\alpha_W(m_Z)= g_4^2/4\pi = 0.0338$ 
and $z_L= 10^{15}$ ($kL = 34.5$) given,  one finds that 
$k=4.7 \times 10^{17} \,$GeV, $m_\KK = 1.48\,$TeV,
$c=0.429$, and $m_H = 49.9 \,$GeV.
The numbers are tabulated for various values of $z_L$ in Table.\ \ref{mH-table}.

\ignore{
With $z_L= 10^{17}$, the values become
$k=5.0 \times 10^{19} \,$GeV, $m_\KK = 1.58\,$TeV,
$c=0.435$, and $m_H=54.4 \,$GeV.  
With $z_L= 10^{10}$, one finds that 
$k=3.9 \times 10^{12} \,$GeV, $m_\KK = 1.21\,$TeV,
$c=0.384$, and $m_H=40.6 \,$GeV.
}

\begin{table}[b,t]
\begin{center}
\begin{tabular}{|c||c|c|c|c|} 
\noalign{\kern 15pt}
\hline
$z_L= e^{kL}$ & $k$ (GeV) & $m_\KK$ (TeV) & $c$ & $m_H$  \\ \hline 
~$10^{17}$~ & ~$5.0 \times 10^{19}$~  & 1.58 & ~0.438~ & ~ 53.5 ~ \\ \hline
~$10^{15}$~ & $4.7 \times 10^{17}$   & 1.48 & 0.429 & ~ 49.9 ~ \\ \hline
~$10^{13}$~ & $4.4 \times 10^{15}$   & 1.38 & 0.417 & ~ 46.1 ~ \\ \hline
~$10^{10}$~ & $3.9 \times 10^{12}$   & 1.21 & 0.388 & ~ 39.9 ~ \\ \hline
~$10^{5}$~ & $2.7 \times 10^{7}$   & 0.86 & 0.226 & ~ 26.9 ~ \\ \hline
~$9.4 \times 10^{3}$~ & $2.3 \times 10^{6}$   & 0.76 & 0. & ~ 23.5 ~ \\ \hline
\end{tabular}
\end{center}
\caption{The Higgs mass $m_H$.  With the value of $z_L$ given, 
$k$, $m_\KK$, $c$ and $m_H$ are determined.   
Input parameters are $m_W = 80.4\,$GeV, 
$\alpha_W = 0.0338$ and $m_t = 172 \,$GeV.
For  $z_L < 9.4 \times 10^{3}$ there is no value for $c$ which reproduces $m_t$.
For $z_L < 905$, $V_\eff(\theta_H)$ with $c=0$  is minimized at $\theta_H = 0, \pi$ so that 
the electroweak symmetry remains unbroken.
}
\label{mH-table}
\end{table}

With $m_W$, $m_t$, $m_b$, $\alpha_W$ and $z_L= e^{kL}$ given, all other
relevant parameters at low energies are determined.  The effective potential is 
minimized at $\theta_H = \pm \half \pi$ where EW symmetry spontaneously
breaks down.
We stress that the Higgs mass $m_H$ is mostly determined by $m_W$, 
$\alpha_W$  and $m_t$. 

It is seen that the Higgs mass is predicted around 50 GeV
for $k = 10^{15} \sim 10^{19} \,$GeV.  One might wonder
if this is in conflict with the  LEP2 bound for $m_H$ which states that
$m_H < 114 \,$GeV is excluded.  We contend that $m_H \sim 50 \,$GeV is 
in no conflict with the LEP2 bound in the current gauge-Higgs unification
scenario.

The crucial observation is that the $ZZH$ coupling vanishes at 
$\theta_H = \onehalf\pi$ as shown in refs.\ \cite{SH1} and \cite{HS2}.
The $WWH$ and $ZZH$ couplings in the $SO(5) \times U(1)_X$ model 
are suppressed, compared with those in the standard model, by a factor
$\cos \theta_H$.  The process $e^+ e^- \go Z \go Z H$ cannot take place 
at $\theta_H = \pm \onehalf \pi$ so that the LEP2 bound is not applicable.  
The $ZZHH$ coupling, on the other hand, 
is multiplied by a factor $\cos 2 \theta_H$ to the coupling  in the 
standard model \cite{Sakamura1} so that  $e^+ e^-  \go ZH H$
can proceed.  Light Higgs particles might have been already produced.
It is of great interest that a similar scenario emerges in a version of
the Minimal Supersymmetric Standard Model (MSSM) where the lightest Higgs 
boson has a different coupling to $Z$ from that of the Higgs boson in the 
standard model~\cite{Kane:2004tk}--\cite{Tobe}\footnote{Note that, 
since the lightest Higgs is not standard model-like, the naive decoupling limit 
cannot be taken in this light Higgs MSSM scenario. In our case, the Kaluza-Klein 
scale $m_\text{KK}$ is related to $m_W$ by Eq.~\eqref{Wmass} so that one 
cannot  arbitrarily take $m_\text{KK}\to\infty$ limit for decoupling.} and 
in the strongly interacting light Higgs scenario~\cite{Giudice1}.
A distinctive feature in the gauge-Higgs unification scenario is that
the light Higgs particle with vanishing $WWH$ and $ZZH$ couplings follows from
the dynamics in the theory, but not by tuning parameters.

We would like to mention that the Higgs mass is expected to remain finite
to all orders in perturbation theory.  It is finite at the one loop level as 
the $\theta_H$-dependent part of the effective potential $V_\eff $ 
is finite as shown originally  in ref.\ \cite{YH1}, generally in ref.\ \cite{YHscgt2} 
and also in the present paper.  The finiteness has been shown at the two loop level
in a toy model of five-dimensional QED.\cite{YHfinite2}

A few comments are in order about the estimate of the Higgs mass given in 
ref.\ \cite{HM}.  It has been argued there, 
without either specifying the detailed fermion content  or performing
explicit computation of the effective potential $V_\eff(\theta_H)$,  
that in generic gauge-Higgs unification models in the RS space 
the Higgs mass should turn out in the range 140 - 280 GeV.
In the present model we have found $m_H \sim 50\,$GeV.  
The discrepancy stems from a couple of sources.  First, in the 
evaluation of the effective potential we observed that the contribution
from the top quark is halved due to the brane mass interactions.  
Second, we found that the effective potential takes the minimum at 
$\theta_H =\onehalf \pi$ whereas $\theta_H = (0.2 \sim 0.3) \pi$
was supposed in ref.\ \cite{HM}.  In the current model 
$m_H \propto m_W/|\sin \theta_H |$ so that smaller $\theta_H$
would give larger $m_H$.  Thirdly, the $c$-dependence of $V_\eff(\theta_H)$
was not well appreciated in ref.\ \cite{HM}.   We have seen that 
for $c \sim 0.43$ there is partial cancellation between contributions from
the top quark and gauge fields.  If $c \sim 0.4$ ($m_t \sim 200\,$GeV), 
then $m_H$ would be increased by 40\% to $73\,$GeV.    
The LEP2 bound $m_H \sim 114\,$GeV would be 
achieved if one takes an  unrealistic value  $m_t \sim 262\,$GeV ($c \sim 0.31$).\footnote{Recall that our model does not need to give $m_H>114\,\text{GeV}$ because of the vanishing $ZZH$ coupling. }
The appearance of the
enhancement factor $kL/2$ in various physical quantities remains valid.

\section{Summary and discussions}

In the present paper we constructed an $SO(5) \times U(1)_X$ 
gauge-Higgs unification model in the RS space with top and bottom quarks
realized in two multiplets in the vector representation ({\bf 5}) of $SO(5)$.
Additional brane fermions are introduced on the Planck brane to make all 
unwanted exotic particles heavy by brane mass terms, and at the same time
to give a bottom quark a finite mass.
Everything follows from equations of motion derived from
the action principle with the orbifold boundary conditions.
The effective change of boundary conditions results for low-lying modes 
of the Kaluza-Klein towers of exotic particles.  The effective potential for
the Wilson line phase and the Higgs mass are  determined from the other
observed quantities.

It was shown that the presence of a top quark triggers the electroweak 
symmetry breaking by the Hosotani mechanism.  The effective potential
was minimized at the Wilson line phase $\theta_H = \pm \onehalf \pi$.
The Higgs mass $m_H$ is predicted, once $m_W$, $\alpha_W$, $m_t$ and
$z_L$ are given. 
 It is found that $m_H \sim 50\,$GeV for $z_L = 10^{15} \sim 10^{17}$.  
The $WWH$ and $ZZH$  couplings vanish at $\theta_H = \pm \onehalf \pi$
so that the LEP2 bound is evaded.   
We stress that the prediction is robust.  It does not depend on the values of
brane masses so long as the scale of the brane masses is much larger than
$m_\KK$.  In short,  the top mass determines the Higgs mass.

One may wonder if the vanishing, or suppression, of the $WWH$ and $ZZH$  
couplings leads to the violation of the tree unitarity in the scattering of 
longitudinal components of $W$ and $Z$.   In  ref.\  \cite{Falkowski2} it has
been shown that  KK excited states of $W$ and $Z$ contribute to
restore the unitarity at high energies through $WW^{(n)}H$ and $ZZ^{(n)}H$ 
couplings.  

Phenomenology of the  Higgs particle is of great interest.  From the study of the 
$SU(3)$ model \cite{HNSS}
it is  expected that Yukawa couplings of the Higgs particle to quarks
are suppressed compared with those in the standard model.
The suppression would be milder for the top quark with $c \sim 0.4$ than
that for lighter quarks with $c > 0.6$.  
The suppressed Yukawa coupling to the bottom quark implies that
the Higgs particle has a rather narrow decay width.

When $\theta_H$ becomes large,  generically large corrections are expected
for the electroweak precision measurements, especially to the $S$ and $T$ 
parameters.\cite{Agashe4, Agashe2, Agashe3, Contino3, Carena2}
Our model,  unlike the preceding ones, does not need any brane dynamics 
for the effective change of the boundary conditions at the TeV brane.
It is manifest that 
our model fits into the criteria of ref.\ \cite{Agashe-custodial} for suppressing
radiative corrections to the $\rho$ ($T$) parameter and $Z b \bar b$ coupling
thanks to the custodial  symmetry in the bulk and on the TeV brane
and the extended $SO(5) \times Z_2 \simeq O(5)$ symmetry.
In ref.\ \cite{Wagner1} it has been pointed out that sizable loop corrections to $T$ 
may result when $t_L$ and $t'_R$ are placed in  one multiplet.
It is important to study such corrections in more detail in our  framework.

The gauge-Higgs unification scenario predicts significant departure from the
standard model, particularly in the Higgs sector.   The forthcoming experiments
at LHC will give us clues in understanding the structure of the symmetry breaking
and the origin of the Higgs particle.

\vskip .5cm

\leftline{\bf Acknowledgments}
The authors would like to thank S.\ Kanemura for an invaluable comment
concerning the LEP2 bound for the Higgs mass.  They are also grateful to
K.\ Agashe and T.\ Takeuchi for enlightening and helpful comments 
on the implications to the EW precision measurements.  
This work was supported in part 
by  Scientific Grants from the Ministry of Education and Science, 
Grant No.\ 20244028, Grant No.\ 20025004 (Y.H. and K.O.), 
Grant No.\ 50324744 (Y.H.),
and Grant No.\ 19740171 (K.O.),
and by Special Postdoctoral Researchers Program at RIKEN (Y.S.).

\def\jnl#1#2#3#4{{#1}{\bf #2} (#4) #3}

\def\Zphys{{\em Z.\ Phys.} }
\def\jssc{{\em J.\ Solid State Chem.\ }}
\def\jpsJ{{\em J.\ Phys.\ Soc.\ Japan }}
\def\ptps{{\em Prog.\ Theoret.\ Phys.\ Suppl.\ }}
\def\PTP{{\em Prog.\ Theoret.\ Phys.\  }}

\def\JMP{{\em J. Math.\ Phys.} }
\def\NPB{{\em Nucl.\ Phys.} B}
\def\NP{{\em Nucl.\ Phys.} }
\def\PLB{{\em Phys.\ Lett.} B}
\def\PL{{\em Phys.\ Lett.} }
\def\PRL{\em Phys.\ Rev.\ Lett. }
\def\PRB{{\em Phys.\ Rev.} B}
\def\PRD{{\em Phys.\ Rev.} D}
\def\PRe{{\em Phys.\ Rep.} }
\def\AP{{\em Ann.\ Phys.\ (N.Y.)} }
\def\RMP{{\em Rev.\ Mod.\ Phys.} }
\def\ZPC{{\em Z.\ Phys.} C}
\def\SCI{\em Science}
\def\CMP{\em Comm.\ Math.\ Phys. }
\def\MPLA{{\em Mod.\ Phys.\ Lett.} A}
\def\IJMPA{{\em Int.\ J.\ Mod.\ Phys.} A}
\def\IJMPB{{\em Int.\ J.\ Mod.\ Phys.} B}
\def\EPJC{{\em Eur.\ Phys.\ J.} C}
\def\PR{{\em Phys.\ Rev.} }
\def\JHEP{{\em JHEP} }
\def\cmp{{\em Com.\ Math.\ Phys.}}
\def\JPA{{\em J.\  Phys.} A}
\def\JPG{{\em J.\  Phys.} G}
\def\NJP{{\em New.\ J.\  Phys.} }
\def\CQG{\em Class.\ Quant.\ Grav. }
\def\ATMP{{\em Adv.\ Theoret.\ Math.\ Phys.} }
\def\ibid{{\em ibid.} }

\renewenvironment{thebibliography}[1]
         {\begin{list}{[$\,$\arabic{enumi}$\,$]}  
         {\usecounter{enumi}\setlength{\parsep}{0pt}
          \setlength{\itemsep}{0pt}  \renewcommand{\baselinestretch}{1.2}
          \settowidth
         {\labelwidth}{#1 ~ ~}\sloppy}}{\end{list}}

\def\reftitle#1{{\it ``#1'' }}    


\vskip 1.cm

\begin{thebibliography}{99}
\small
\baselineskip=14pt

\leftline{\bf References}


\bibitem{review1}
H.~C.~Cheng,
arXiv:0710.3407 [hep-ph].
  \reftitle{Little Higgs, Non-standard Higgs, No Higgs and All That}

\bibitem{TASI1}
C.\ Csaki, J.\ Hubisz and P.\ Meade,
arXiv: hep-ph/0510275.
\reftitle{TASI Lectures on Electroweak Symmetry Breaking from Extra Dimensions}


\bibitem{YH1}
Y.\ Hosotani, \jnl{\PLB}{126}{309}{1983}.
\reftitle{Dynamical Mass Generation by Compact Extra Dimensions}

\bibitem{YH2}
Y.\ Hosotani, \jnl{\AP}{190}{233}{1989}.
\reftitle{Dynamics of Nonintegrable Phases and Gauge Symmetry Breaking}

\bibitem{HHHK}
N.\ Haba, M.\ Harada, Y.\ Hosotani and Y.\ Kawamura, 
\jnl{\NPB}{657}{169}{2003};   
{\it Erratum}, {\it ibid.}  B{\bf 669} (2003) {381}.
\reftitle{Dynamical Rearrangement of Gauge Symmetry on the Orbifold $S^1/Z_2$}


\bibitem{Pomarol1}
A.\ Pomarol and M.\ Quiros, \jnl{\PLB}{438}{255}{1998};
\reftitle{The Standard Model from extra dimensions}


\bibitem{Antoniadis1}
I.\ Antoniadis, K.\ Benakli and M.\ Quiros,
\jnl{\it New. J.\ Phys.}{3}{20}{2001}.
\reftitle{Finite Higgs mass without Supersymmetry}

\bibitem{Lim1}
M.\ Kubo, C.S.\ Lim and H.\ Yamashita,
 \jnl{\MPLA}{17}{2249}{2002}.
\reftitle{The Hosotani Mechanism in Bulk Gauge Theories with an Orbifold Extra   Space $S^1/Z_2$}

\bibitem{Scrucca}
C.A.\ Scrucca, M.\ Serone, and L.\ Silvestrini,
\jnl{\NPB}{669}{128}{2003}.
\reftitle{Electroweak symmetry breaking and fermion masses from extra dimensions}

\bibitem{Nomura1}
G.\ Burdman and Y.\ Nomura, \jnl{\NPB}{656}{3}{2003}; 
\reftitle{Unification of Higgs and Gauge Fields in Five Dimensions}

\bibitem{Csaki1}
C.\ Csaki, C.\ Grojean and H.\ Murayama, \jnl{\PRD}{67}{085012}{2003};
\reftitle{Standard Model Higgs From Higher Dimensional Gauge Fields}

\bibitem{HHKY}
N.\ Haba,  Y.\ Hosotani,  Y.\ Kawamura and T.\ Yamashita, 
\jnl{\PRD}{70}{015010}{2004};
\reftitle{Dynamical symmetry breaking in Gauge-Higgs unification on orbifold}


\bibitem{HNT2}
Y.\ Hosotani, S.\ Noda and K.\ Takenaga,
\jnl{\PLB}{607}{276}{2005}.
\reftitle{Dynamical Gauge-Higgs Unification in the Electroweak Theory}

\bibitem{Csaki2}
G.\ Cacciapaglia, C.\ Csaki and S.C.\ Park,
\jnl{\JHEP}{0603}{099}{2006}.
\reftitle{Fully Radiative Electroweak Symmetry Breaking}

\bibitem{Panico1}
G.\ Panico, M.\ Serone and A.\ Wulzer, 
\jnl{\NPB}{739}{186}{2006};
\reftitle{A Model of electroweak symmetry breaking from a fifth dimension}
\jnl{\NPB}{762}{189}{2007}.
\reftitle{Electroweak Symmetry Breaking and Precision Tests with a Fifth Dimension}


\ignore{
C.A.\ Scrucca, M.\ Serone, L.\ Silvestrini and A.\ Wulzer,
\jnl{\JHEP}{0402}{49}{2004};
\reftitle{Gauge-Higgs Unification in Orbifold Models} 
}



\bibitem{Lim2}
H.\ Hatanaka, T.\ Inami and C.S.\ Lim, 
\jnl{\MPLA}{13}{2601}{1998}.
\reftitle{The Gauge Hierarchy Problem and Higher Dimensional Gauge Theories}


\bibitem{Pomarol2}
R.\ Contino, Y.\ Nomura and A.\ Pomarol, \jnl{\NPB}{671}{148}{2003}.
\reftitle{Higgs as a holographic pseudo-Goldstone boson}

\bibitem{Agashe4}
K.\ Agashe, A.\ Delgado, M.J.\ May and R.\ Sundrum,
\jnl{\JHEP}{0308}{050}{2003}.  
\reftitle{RS1, custodial isospin and precision tests}

\bibitem{Agashe2}
K.\ Agashe, R.\ Contino and A.\ Pomarol, 
\jnl{\NPB}{719}{165}{2005}.
\reftitle{The minimal composite Higgs model}


\bibitem{Agashe3}
K.\ Agashe and R.\ Contino, 
\jnl{\NPB}{742}{59}{2006}.
\reftitle{The minimal composite Higgs model and electroweak precision tests}


\bibitem{HM}
Y.\ Hosotani and M.\ Mabe, \jnl{\PLB}{615}{257}{2005}.
\reftitle{Higgs boson mass and electroweak-gravity hierarchy
from dynamical gauge-Higgs unification in the warped spacetime}


\bibitem{Oda1}
K.\ Oda and A.\ Weiler, \jnl{\PLB}{606}{408}{2005}.
\reftitle{Wilson Lines in Warped Space: Dynamical Symmetry Breaking and Restoration}

\bibitem{HNSS}
Y.\ Hosotani, S.\ Noda, Y.\ Sakamura and S.\ Shimasaki, 
\jnl{\PRD}{73}{096006}{2006}.
\reftitle{Gauge-Higgs Unification and Quark-Lepton Phenomenology 
in the Warped Spacetime}



\bibitem{Carena}
M.\ Carena, E.\ Ponton, J.\ Santiago and C.E.M.\ Wagner,
\jnl{\NPB}{759}{202}{2006}.   
\reftitle{Light Kaluza Klein States in Randall-Sundrum Models with Custodial SU(2)}

\bibitem{Contino3}
R.\ Contino, L.\ Da Rold and A.\ Pomarol,
\jnl{\PRD}{75}{055014}{2007}.   
\reftitle{Light custodians in natural composite Higgs models}


\bibitem{Carena2}
M.\ Carena, E.\ Ponton, J.\ Santiago and C.E.M.\ Wagner,
\jnl{\PRD}{76}{035006}{2007}.
\reftitle{Electroweak constrains on warped models with custodial symmetry}

\bibitem{SH1} Y.\ Sakamura and Y.\ Hosotani, \jnl{\PLB}{645}{442}{2007}. 
\reftitle{WWZ, WWH and ZZH couplings in the dynamical gauge-Higgs unification
in the warped spacetime}

\bibitem{HS2} Y.\ Hosotani and Y.\ Sakamura, \jnl{\PTP}{118}{935}{2007}. 
\reftitle{Anomalous Higgs couplings in the $SO(5) \times U(1)$ gauge-Higgs 
unification in warped spacetime}

\bibitem{Falkowski1}
A.\ Falkowski, \jnl{\PRD}{75}{025017}{2007}.
\reftitle{Holographic pseudo-Goldstone boson}

\bibitem{Panico2}
G.\ Panico and A.\ Wulzer, 
 \jnl{\JHEP}{0705}{060}{2007}.
 \reftitle{Effective action and holography in 5D gauge theories}

\bibitem{Sakamura1}
Y.\ Sakamura,  \jnl{\PRD}{76}{065002}{2007} (arXiv:0705.1334 [hep-ph]).
\reftitle{Effective theories of gauge-Higgs unification models in warped spacetime}

\bibitem{Wagner1}
A.D.\ Medina,  N.R.\ Shah and C.E.M.\ Wagner,
 \jnl{\PRD}{76}{095010}{2007} (arXiv:0706.1281 [hep-ph]).
 \reftitle{Gauge-Higgs Unification and Radiative Electroweak Symmetry 
 Breaking in Warped Extra Dimensions}

\bibitem{Wagner2}
M.\ Carena, A.D.\ Medina, B.\ Panes, N.R.\ Shah and C.E.M.\ Wagner,
arXiv:0712.0095 [hep-ph]
\reftitle{Collider phenomenology of gauge-Higgs unification scenarios 
in warped extra dimensions}


\bibitem{Falkowski2}
A.\ Falkowski,  S.\ Pokorski and J.P.\ Roberts, 
\jnl{\JHEP}{0712}{063}{2007} (arXiv:0705.4653 [hep-ph]).
\reftitle{Modelling strong interactions and longitudinally polarized 
vector boson scattering}

\bibitem{Falkowski3}
A.\ Falkowski,  \jnl{\PRD}{77}{055018}{2008} (arXiv:0711.0828 [hep-ph]). 
\reftitle{Pseudo-goldstone Higgs production via gluon fusion}

\bibitem{Hatanaka1}
H.\ Hatanaka,  arXiv: 0712.1334 [hep-th].
\reftitle{Radiatively Induced Spontaneous Symmetry Breaking by Wilson Line 
in a Warped Extra Dimension}

\bibitem{Panico3}
G.\ Panico, E.\ Pronton, J.\ Santiago and M.\ Serone,  arXiv:0801.1645 [hep-ph].
\reftitle{Dark Matter and Electroweak Symmetry Breaking in Models 
with Warped Extra Dimensions}

\bibitem{Yamashita}
N.\ Haba, S.\ Matsumoto, N.\ Okada and T.\ Yamashita, arXiv:0802.3431 [hep-ph]
\reftitle{Effective Potential of Higgs Field in Warped Gauge-Higgs Unification}

\bibitem{Csaki3}
C.\ Csaki, A.\ Falkowski and A.\ Weiler,  arXiv:0804.1954 [hep-ph].
\reftitle{The Flavor of the Composite Pseudo-Goldstone Higgs}


\bibitem{Giudice1}
G.F.\ Giudice, C.\ Grojean, A.\ Pomarol and R.\ Rattazzi,
\jnl{\JHEP}{0706}{045}{2007}.
\reftitle{The Strongly Interacting Light Higgs}


\bibitem{Lim3}
C.S.\ Lim and N.\ Maru, hep-ph/0703017. 
\reftitle{Calculable One-Loop Contributions to S and T Parameters 
in the Gauge-Higgs Unification}

\bibitem{Lim4}
Y.\ Adachi, C.S.\ Lim and N.\ Maru,  
\jnl{\PRD}{76}{075009}{2007} (arXiv:0707.1735 [hep-ph]).
\reftitle{Finite anomalous magnetic moment in the gauge-Higgs unification}


\bibitem{Agashe-custodial}
K.\ Agashe, R.\ Contino, L.\ Da Rold and A.\ Pomarol, 
\jnl{\PLB}{641}{62}{2006}.
\reftitle{A custodial symmetry for $Z b \bar b$}




\bibitem{RS1}
L.\ Randall and R.\ Sundrum,  \jnl{\PRL}{83}{3370}{1999}.
\reftitle{A Large mass hierarchy from a small extra dimension}

\bibitem{GP}
T.\ Gherghetta and A.\ Pomarol,
\jnl{\NPB}{586}{141}{2000}.
\reftitle{Bulk fields and supersymmetry in a slice of AdS}



\bibitem{Arkani}
N.\ Arkani-Hamed and M.\ Schmaltz,
\jnl{\PRD}{61}{033005}{2000}.
\reftitle{Hierarchies without symmetries from extra dimensions}


\bibitem{Kane:2004tk}
  G.~L.~Kane, T.~T.~Wang, B.~D.~Nelson and L.~T.~Wang,
  \jnl{\PRD}{71}{035006}{2005}.
\reftitle{Theoretical implications of the LEP Higgs search}

\bibitem{Drees}
M.\ Drees, 
\jnl{\PRD}{71}{115006}{2005}.
\reftitle{A Supersymmetric Explanation of the Excess of Higgs-Like Events at LEP}

\bibitem{Kim:2006mb}
  S.~G.~Kim, N.~Maekawa, A.~Matsuzaki, K.~Sakurai, A.~I.~Sanda and T.~Yoshikawa,
  \jnl{\PRD}{74}{115016}{2006}.
\reftitle{A Solution for Little Hierarchy Problem and $b \go s \gamma$}

\bibitem{Tobe}
A.\ Belyaev, Q.H.\ Cao,  D.\ Nomura, K.\ Tobe, C.-P. Yuan,
\jnl{\PRL}{100}{061801}{2008}.
\reftitle{Light MSSM Higgs boson scenario and its test at hadron colliders}

\bibitem{YHscgt2}
Y.\ Hosotani, in the Proceedings of 
{\it ``Dynamical Symmetry Breaking"},  ed. M. Harada and K. Yamawaki 
(Nagoya University, 2004), p.\ 17. (hep-ph/0504272).
\reftitle{Dynamical Gauge Symmetry Breaking by Wilson Lines
in the Electroweak Theory}

\bibitem{YHfinite2}
Y.\ Hosotani, N.\ Maru, K.\ Takenaga and T.\ Yamashita, 
\jnl{\PTP}{118}{1053}{2007}.
\reftitle{Two Loop finiteness of Higgs mass and potential in the gauge-Higgs 
unification}







\end{thebibliography}
\end{document}